\documentclass[manuscript]{aastex}

\shorttitle{Detecting Extrasolar Planets with Integral Field Spectroscopy}
\shortauthors{Berton A., Gratton R.G., Feldt M., et al.}

\begin{document}


\title{Detecting Extrasolar Planets with Integral Field Spectroscopy}


\author{A. Berton\altaffilmark{1}, R. G. Gratton\altaffilmark{2}, M. Feldt\altaffilmark{1}, T. Henning\altaffilmark{1}, S. Desidera\altaffilmark{2}, M. Turatto\altaffilmark{2}, \break H. M. Schmid\altaffilmark{3} and R. Waters\altaffilmark{4}}

\email{berton@mpia.de, gratton@pd.astro.it, feldt@mpia.de, henning@mpia.de, desidera@pd.astro.it, turatto@pd.astro.it, schmid@astro.phys.ethz.ch, rensw@science.uva.nl}




\altaffiltext{1}{Max-Planck-Institut f\"ur Astronomie (MPIA), K\"onigstuhl 17, D-69117, Heidelberg, Germany}
\altaffiltext{2}{Osservatorio Astronomico di Padova INAF, Vic. Osservatorio 5, I-35122, Padova, Italy}
\altaffiltext{3}{Institute of Astronomy, ETH Zentrum, 8092 Zurich, Switzerland}
\altaffiltext{4}{Astronomical Institute, University of Amsterdam, Kruislaan 403, 1098 SJ Amsterdam, The Netherlands}


\begin{abstract}
Observations of extrasolar planets using Integral
Field Spectroscopy (IFS), if coupled with an extreme Adaptive Optics system and
analyzed with a Simultaneous Differential Imaging technique (SDI), are
a powerful tool to detect and characterize extrasolar planets directly; they enhance the signal of the planet and, at the same
time, reduces the impact of stellar light and consequently important noise sources like speckles. In order to verify the efficiency of such a technique,
we developed a simulation code
able to test the capabilities of this
IFS-SDI technique for different kinds of planets and telescopes, modelling the atmospheric and instrumental noise sources.
The first results obtained by the simulations show that many
significant extrasolar planet detections are indeed possible using the present 8m-class 
telescopes within a few hours of exposure time. The procedure adopted to simulate IFS 
observations is presented here in detail, explaining in particular how we 
obtain estimates of the speckle noise, Adaptive Optics corrections, specific
instrumental features, and how we test the efficiency of the SDI technique to increase the signal-to-noise ratio of the planet detection. The most
important results achieved by simulations of various objects, from 1\,M$_J$ to
brown dwarfs of 30\,M$_J$, for observations with an 8 meter telescope, are then 
presented and discussed.
\end{abstract}





\section{Introduction}

At present, around 150 extrasolar planets have been found around stars
other than the Sun\footnote{See for example the continuously updated Extrasolar Planet Encyclopedia web site (\url{www.obspm.fr/encycl/encycl.html}) and references to published papers therein.}. Nearly all of them have been detected in indirect ways,
generally by observing the effects of gravitational perturbations of the host star due to
the presence of the companion. These perturbations can be
measured by radial velocity changes or precise astrometry. 
In addition, transits and microlensing have been successfully applied to detect planets.
Direct detections of extrasolar planets, with the present technology available,
are still a challenging goal to achieve. Only recently the first
observations of the signals coming directly from planetary mass objects have been possible,
in some particular cases of young companions around low-mass stars or brown dwarfs
\citep[for example see][]{cha}; but in general the detection of Jupiter-like 
planets around Solar-type stars remains a challenge.\\
Difficulties in direct detection of faint companions arise from the complex 
structure of
the stellar Point Spread Function (PSF) which swamps the light of the
planet. This stellar halo is usually brighter than the planet at every 
wavelength, and it is also a source of noise, in particular the speckle noise.\\
In Sec.\ref{sec:speckle} it is explained why atmospheric speckles are the most important
source of noise that we have to consider in direct detections of planets. 
Thus,
an efficient method for speckle noise reduction is required. Here we suggest
a Simultaneous Differential Imaging technique (SDI) applied to
IFS observations of the star-planet system, as described in Sec.\ref{sec:sdi}.
In order to test the capability of such a technique in reducing speckle noise 
and detecting planets, we developed a dedicated piece of software, presented in
Sec.\ref{sec:code}, that aims to simulate as accurately as possible an 
observation
of an extrasolar planet using an IFS. From the analysis of the results of the
simulations, it has been possible to understand which kinds of planets are
really detectable and characterizable for various telescope diameters, e.g.
an 8m-class telescope. Some encouraging results of these simulations are 
presented in Sec.\ref{sec:results}.

\section{The speckle noise}
\label{sec:speckle}

For ground-based observations, speckles are mainly due to atmospheric turbulence and they result from the interference of light from many coherent patches of
typical diameter $r_0$ (called the ``Fried radius''), distributed over the
full aperture of the telescope. A detailed description of the formation of
speckles can be found in \citet{rac}. Speckles look like bright spots of
typical Full Width Half Maximum (FWHM)
$\sim \lambda / D$ radians, where $\lambda$ is the wavelength and $D$ is the diameter
of the telescope aperture. The number of speckles per unit area, in case of
$D \gg r_0$ \citep{rod} is:

\begin{equation}
\label{eq:ns}
n_s \simeq \frac{0.342}{\pi \left( \lambda / D \right) ^2} .
\end{equation}

\noindent This means that the larger the diameter of the telescope, the 
higher the number of speckles.\\
The spatial distribution of speckles, the so-called speckle pattern, is a
direct consequence of the wavefront corrugation by atmospheric turbulence and
optical imperfections. Hereafter we will focus our attention on speckles due
to the atmosphere. 
The atmospheric speckle lifetime, called ``coherence time'' $\tau_0$, is proportional to
$r_0/\Delta v$, where $\Delta v$ is the velocity dispersion in the turbulent
seeing layers across the telescope line of sight. This speckle variability,
related to the variability of the wavefront corrugation, results in ``speckle boiling'' and is the source of
speckle noise.

The statistics of speckle noise variance are not very simple, because the speckles are
not independent events by the nature of the interference that produces them.
In the presence of an Adaptive Optics system (AO) for compensating atmospheric turbulence, 
\citet{rac} found, for the speckle variance in the PSF
halo, the expression:

\begin{equation}
var(speckles) \simeq 1.7 \pi \left[ \left( \frac{\lambda}{D} \right) ^2 
\left( 1-S \right) F_* f_{halo}(\theta) \right] ^2 \frac{\tau_0}{t},
\label{eq:var}
\end{equation}

\noindent where $F_*$ is the total flux from the star, $f_{halo}$ indicates the
radial profile of the PSF halo ($\theta$ is the angular distance from the
center), $t$ is the exposure time and $S$ is the Strehl ratio given by the AO
compensation. Since $F_*$ is proportional to the exposure time $t$, this
equation shows that the variance of speckles is proportional to $t$. This
variance can actually be reduced in case of an AO system able to reach very high
Strehl ratios $S$.\\
Considering the variance of the photon noise of the PSF halo, which is

\begin{equation}
var(photons) = \pi \left( \frac{\lambda}{D} \right) ^2 \left( 1-S \right)
F_* f_{halo}(\theta),
\end{equation}

\noindent a comparison between the two variances shows that the speckle
one is usually much larger than the photonic one.\footnote{A separate discussion has to be done for long-time speckles, or
``super-speckles'', which are more related to static aberrations than to the
atmosphere. It is not easy to determine their life time, that will be much
longer than $\tau_0$: in the worst case, they don't change during all the
integration time.} 

We can conclude that the speckles are the most important source of noise that
we have to face searching directly for extrasolar planets from the ground, and we must find a way for reducing them. Obviously, the best solution would be
space telescopes, outside the terrestrial atmosphere: see e.g. the projects
DARWIN \citep{fri} and Terrestrial Planet Finder \citep[TPF,][]{bei}; even so,
because of the complexity of a space mission, to exploit new ways for
speckle noise reduction for ground-based observations is not a worthless effort.  
From equation \ref{eq:var}, it is clear that an extreme
AO system is mandatory in order to reduce dramatically the speckle 
intensity: AO systems can reduce seeing disk and get a sufficient spatial resolution for our investigation, and at the same time can improve the value of the Strehl ratio $S$. Present AO systems are able to reach very high Strehl 
ratios (up to 0.7 at $\lambda$=2.2$\mu$m) but they are still far from being perfect ($S$=1). Still after AO correction, speckle noise is orders of magnitude larger than the photon noise; therefore the reduction of the speckle noise to a level comparable to the photon noise can be achieved only using additional techniques. 

Many different methods have been suggested, almost all of
them based on a combination of XAO and coronography. In fact, the coronograph can improve the sensitivity of an imaging system to faint structures surrounding the bright source \citep{siv}. In addition, noise can be reduced using a narrow-band coronographic
interferometer \citep{cod}, nulling interferometry \citep{bra,woo}, adaptive holograms for removing
coronographic residues \citep{lab} or synchronous 
interferometric speckle subtraction \citep{guy}. The polarization of the
light reflected by the planet can be also used to reduce the effect of
speckles \citep{bab,sta}.
A large improvement in planet detection, in particular for Earth-like planets,
can be reached by a combination of AO, coronography and Extremely Large
Telescopes (ELTs) because of the large amount of photons collectible in short
exposures \citep{che,yai}.

\section{The IFS-Simultaneous Differential Imaging technique}
\label{sec:sdi}

The main idea behind SDI is to use differential measurements in two bands, only one of them containing the planet signal, to enhance the 
features that make the planet distinguishable from the central star
\citep[see, e.g.,][]{smi,ros,rac,mar}. Practically this can be done
choosing a couple (or more) wavelengths at which the light of the star presents very small differences in intensity, but at the same time the light of the
planet is changing drastically because of absorption bands\footnote{A similar approach is also possible in polarimetry, exploiting the fact that light reflected by a planet is partially polarized \citep{sta}.}.
The wavelengths must be chosen appropriately for each kind of target. In the case of
low-mass brown dwarfs or gaseous giant planets, for instance, the methane
absorption bands in the near-infrared are the best solution (see Sec.\ref{sec:spettri} for details). The simultaneous
observation of the star-planet system at two wavelengths on and off the methane absorption band, done using
special narrow-band filters e.g. in NACO-SDI on ESO's VLT \citep{len}, gives
a set of images that, after a suitable re-scaling in wavelength, can be 
combined using various algorithms in order to subtract the stellar light
\citep[see e.g.][]{bil1,bil2}.
Because of the simultaneity of the observations, the subtraction of the stellar
light leads also to the subtraction of speckles and the reduction
of the speckle noise. A recent example of a low mass companion (a cool methane
rich brown dwarf) using NACO-SDI is described in \cite{bil3}.

Although SDI was originally proposed for two wavelengths only and for a simple
difference operation, \citet{mar} showed that the speckle noise reduction can be much more
efficient if observations at 3 wavelengths are available, using a 
``double-difference''. Starting from here, it is reasonable to
assume that a larger number of images at different wavelengths, 
taken with a small regular spectral separation, can result in even better reduction of 
speckle noise. This idea suggested the use of an integral field spectrograph
for collecting data simultaneously at a large number of different wavelengths,
given by the total spectral length and the spectral resolution of the 
disperser. IFS is preferable to a single slit spectrograph, because it allows
the detection of the companion even if its position is unknown.  
Moreover, an IFS has some advantages with respect to a filter-based SDI in reducing 
speckle noise and in increasing the signal-to-noise ratio of the detection,
because of the possibility to analyze a larger amount of information at
many different wavelengths, and these advantages are discussed in detail in
Sec.\ref{sec:why}.
There is however a disadvantage to using this technique owing to the instrumental complexity and complicated data reduction procedure.

\subsection{The stellar and planetary spectra}
\label{sec:spettri}

To design an instrument dedicated to the detection of planets using the
IFS-SDI technique, it is necessary to have a clear idea of the spectral
features that will be observed, in order to choose the best spectral region for observations, and to optimize the spectral resolution and spectral coverage of the instrument. In this section, we discuss these aspects of the observation of giant
Jupiter-like planets around solar-type stars.

\begin{figure}
\epsscale{.70}
\plotone{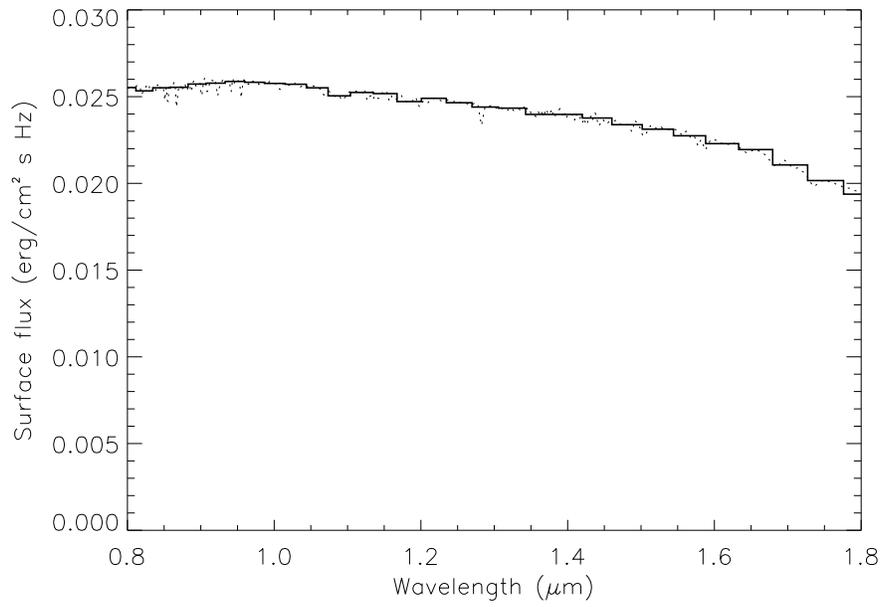}
\caption{Near-Infrared spectrum (between 0.8 and 1.8$\mu$m) of a G2V star at the spectral resolution of 30 (solid line) and 750 (dotted line). In this plot are not included the absorptions caused by terrestrial atmosphere. Data taken from \citet{bur}. \label{fig:star}}
\end{figure}

Models of giant planets, like those of \citet{bur} for irradiated planets and
\citet{bur2} for isolated objects, clearly show
that the most interesting features expected in the spectra are the broad
absorption bands of water and methane in the near-infrared. These bands can be
studied at very low spectral resolution (15 or 30, see 
Fig.\ref{fig:plan}). 
From these plots we can see that the methane dominated spectra are characterized by ``emission peaks'', the portions of spectrum not affected by absorptions, which are expected to be constant in wavelength, but their intensities are expected to change from object to object.
At the same resolution and in the same spectral
region the spectrum of a solar-type star is nearly featureless and may be
fitted by a polynomial curve (see Fig.\ref{fig:star}). The combination of strong
planetary features and very smooth stellar spectra makes this spectral region
optimal for application of the SDI technique, which is sensitive to the variations in contrast between star and planet.


\section{Description of the simulation code}
\label{sec:code}

The simulation code presented in this section is called CSP (CHEOPS Simulation Program) and has been developed in the context
of the CHEOPS project \citep[CHaracterizing Exo-planets by Opto-infrared Polarimetry and Spectroscopy,][]{fel}. This project was born as an answer to ESO's call 
for a ``Planet Finder'' at the VLT. The concept of this
instrument was based on an extreme adaptive optics system combined with two
scientific channels. The first channel is a polarization analyzer based on
ZIMPOL \citep{pov,gis}; the second channel is an IFS, based on the SAURON concept \citep{bac} but adapted
to the diffraction limit case relevant to detection of extrasolar giant planets. This IFS consists of a microlens array composed of
254$\times$254 hexagonal microlenses covering a field of view of 3\farcs5$\times$3\farcs5.
The disperser is an Amici prism, which can give spectra with a nearly constant 
resolution of $R$=15 in the spectral region between 0.9 and 1.7$\mu$m. The
resolution has been chosen in order to allow the investigation of the broad
methane absorption bands, and at the same time to sample all the requested
spectral range with a few pixels on the detector (length of $\sim$20
pixels). This length, the resolution and the microlens pitch (0\farcs0138)
have been chosen in order to allow a very low level of cross-talk between
spectra ($\ge$10$^{-4}$)\footnote{The resolution of 15 has been chosen also to
  optimize the contrast between the spectral features we want to observe
  (molecular bands like methane) and the length of the spectra on the detector.}. It is important to note that in the CHEOPS design we
have only two refractive elements before the IFS, which are an Atmospheric
Dispersion Corrector (ADC), included in the AO system, and a dichroic beam splitter: since these two
elements don't contribute in a relevant way to chromatic aberrations (the
latter has no optical power, the former is composed by flat surfaces), then
the total chromatic error is considered to be negligible. A detailed
description of the CHEOPS IFS can be found in \citet{cla}. 

Starting from the technical definitions of CHEOPS, we developed a code with 
the goal of simulating observations of extrasolar planets using an
extreme AO-assisted IFS,
focusing our attention on the sources of noise affecting these observations.
In particular, an accurate simulation of the dominant speckle noise was required, 
beside all the detector noise sources, instrumental aberrations and so on. 
Simulated observation can then be used to test the efficiency of IFS-SDI for
finding planets and to understand exactly which kind of object can be 
actually detected which such an instrument.

An interesting aspect of this simulation code is the possibility to allow simulations of different kinds of telescopes
and different technical configurations. The code can be used for simulating observations with both an 8 meter
telescope and an ELT. It can also be adapted,
knowing the appropriate spectral features, to the detection of earth-like planets
rather than hot and cold Jupiters. 
In this paper we focus our attention on the results related to the case of an 8 meter telescope and Jupiter-like planets: the case of ELTs and Earth-like planets will be exploited in a forthcoming paper.

A large number of input 
parameters of different nature are required by the simulation code. In order to simulate the AO correction all relevant parameters describing
the atmospheric turbulence must be defined ($r_0$, $\tau_0$, etc.). In addition, the diameter
of the telescope and the optical path of the instrument has to be known as well as
the expected static aberrations. The shape, the focal length and the expected
scattered light given by the microlenses has to be considered. Finally,
the detector features strongly affect the final signal-to-noise ratio.
The output given by the simulation code is a single image, on which all the
spectra produced by the IFS are present, showing the spectra of the star and,
usually hidden, the spectra of a given number of planets.

\subsection{Simulation of atmospheric turbulence and AO correction}
\label{sec:ao}

Data concerning atmospheric turbulence, telescope vibration and
Adaptive Optics correction are elaborated as a set of images, phase
screens, which represents the wavefronts associated with observations
provided by AO. A phase screen of the atmospheric turbulence is
produced defining the spatial properties of the screen by a von
K\'arm\'an spectrum, using a procedure based on \citet{lan}.  A single
layer atmosphere was simulated corresponding to seeing conditions
of 0\farcs6, 0\farcs85 and 1\farcs0 at $\lambda = $550\,nm.  $\tau_0$
was always 0.01\,s. Telescope vibrations were added as time-dependent
tip and tilt terms at 14\,Hz and 48\,Hz with amplitudes typical for the
VLT; an ADC is supposed to be present within the AO system. As this paper is not concerned with the details of the AO system, a
very simple AO simulation technique was adopted where the wavefront
sensor is simulated by re-sampling the input wavefront itself to the
sensor resolution of 40 apertures across the 8.0\,m primary mirror.
This corresponds exactly to the number and geometry of the actuators
in the assumed deformable mirror. The ``sensor'' information was fed
into the control loop which uses a simple integrator controller
running at a loop speed 2\,kHz.  This controller produces another
40$\times$40 matrix which represent the mirror
commands.  Since it is still in the same units as the wavefront
itself, the rest of the system is simulated simply by re-expanding
this 40$\times$40 matrix to the original spatial scale of the
wavefront and subtracting it from the input phase screen appropriate
for the moment in time when the mirror signals are applied.

Simulating AO in this way we do not include specific effects of
certain wavefront sensor types and other specialities, but we do get a
rather optimistic estimate what an XAO system could achieve on an 8\,m
telescope.  The resulting Strehl ratios vary between 0.4 and 0.8.  It
should be noted that with this technique we do see a strong dependence
of planet detectabilities from seeing values (see Sec.\ref{sec:seeing}), but since the AO
simulation is not really detailed, one should rather assume the Strehl
number as the input variable of that dependency.  How the Strehl ratio
depends on seeing has to be determined in detailed AO simulations not
included in this work.  On the output side of the atmosphere-AO
simulation chain, every phase screen represents the wavefront at a
precise moment of the observation and it is related to a very short
interval of time. This interval is shorter than the coherence time of
the turbulence. Each screen differs slightly from the previous one.
Phase screens are monochromatic, but they can easily be converted to
the required wavelength using appropriate scale factors.

\subsection{Creation of the PSF and the speckle patterns}

Given a phase screen and the entrance pupil of the telescope,
the code transforms the phase screen, a measure of phase $\phi$, in an electric field $E=Ae^{i \phi}$,
where the amplitude $A$ is given by the pupil and $\phi$ by the phase screen.
The PSF related to that screen is obtained using a Fourier Transform:

\begin{equation}
psf=FT(E) =\int\!\!\!\int_{-\infty}^{+\infty} 
 E(x,y) {\mathrm{e}}^{-i2\pi \left(f_xx + f_yy \right)} {\mathrm{d}}x {\mathrm{d}}y,
\end{equation}


\noindent where $f_x$ and $f_y$ are the coordinates in frequency space. 

We simulated a simplified model of Lyot
coronograph, in order to avoid the saturation of the central peak. Such a coronograph would also
be useful in reducing the effects of diffraction rings and patterns. It is included by applying 
on the focal plane a central obscuration (with a diameter of 0.55 arcseconds at
the shortest wavelength) covering the central stellar peak,
and then a Lyot mask on the pupil plane. This has been modeled as an aperture
on the pupil plane with transmission equal to 1 within 95$\%$ of the diameter
of the pupil of the telescope, and it is slightly apodized beyond this value
using a gaussian function. All these choices have been totally arbitrary
because the final configuration of the coronograph for IFS has been not
decided yet. When a definitive coronograph will be chosen, we will be able to
improve the simulation of this part of the code. For the moment, we stick with
this simplified model.

This PSF has been obtained from a single phase screen. Therefore, it is
related to a short interval of time and it will be characterized by 
an instantaneous speckle pattern. In order to simulate a longer exposure time we
perform a suitable number of repeats and sum
the resulting PSFs into one.     

Our code convolves this PSF with an ``object'' image that is a set of 
point-like sources, the star and one or more planets, characterized by
different intensities. The intensities are provided as a function of the
wavelength at which the code is working. In this way the simulated PSFs are related
with a specific resolution element of the spectrum.

\begin{figure}
\epsscale{1.0}
\plotone{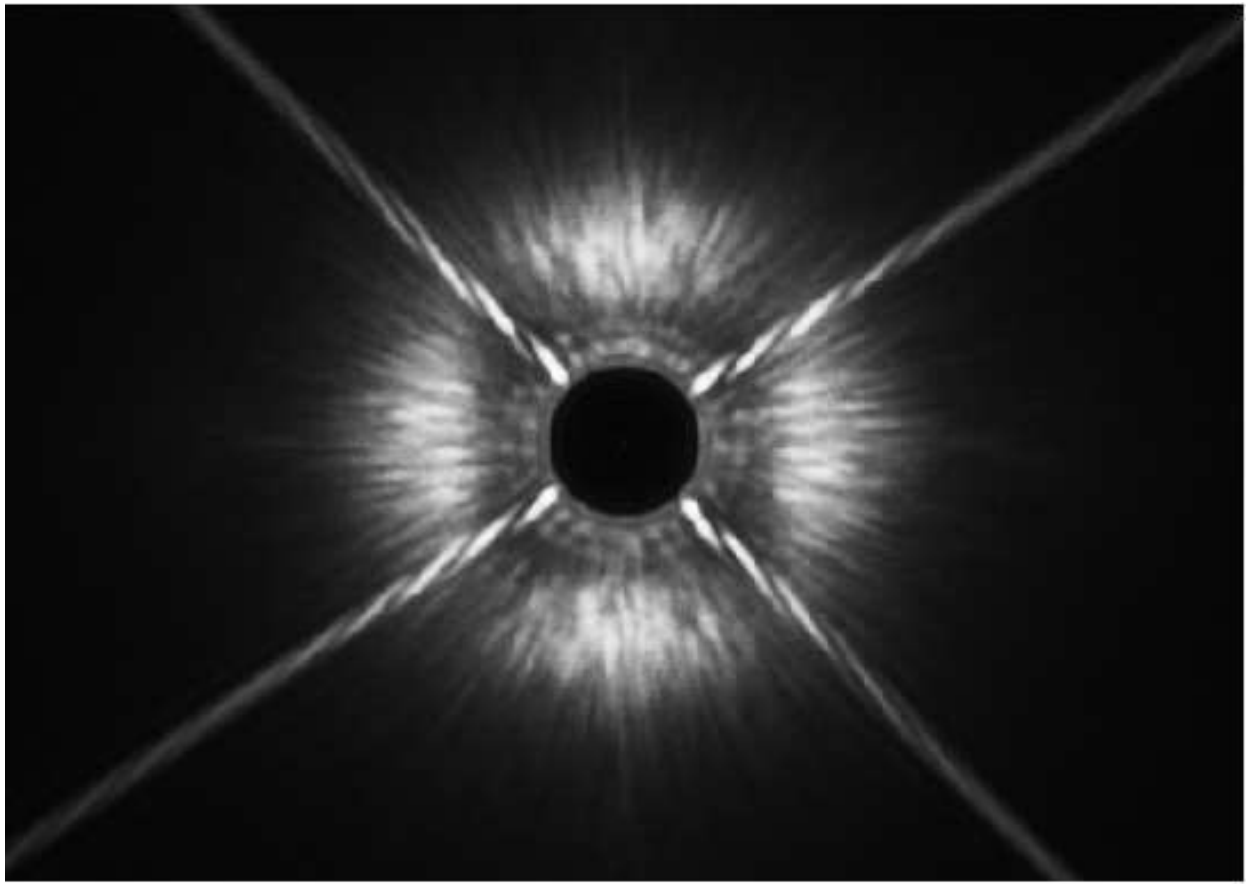}
\caption{Simulated Point Spread Function of a G0 V star at the VLT (prototype of an 8m-class telescope). It corresponds to an integration time of 0.5s, with a Strehl Ratio $\sim$0.8 at $\lambda$=1.0$\mu$m. The width of the simulated resolution element ($\Delta \lambda$) is 0.04$\mu$m. VLT's spiders are clearly visible, as well as the central obscuration with a diameter of 0\farcs55 due to the coronograph. \label{fig:psf}}
\end{figure}

Moreover, every PSF produced in this way is strictly monochromatic. In reality the resolution is very low and the spectral resolution elements are quite
large (with $R$=15 in the near-infrared we expect a $\Delta \lambda \sim$0.04$\mu$m).
We solved this problem in a statistical way exploiting the fact that the typical lifetime of
a phase screen is a fraction of a millisecond, while a simulated exposure time
is usually much larger than 1s. We then calculated the PSF not only at the wavelength $\lambda$, center of the resolution element, but also at  $\lambda \pm \lambda_r$, where
$\lambda_r \le \Delta \lambda / 2$ is a small random value. With this procedure we have
enough different values of $\lambda_r$ to sample all the element uniformly. For this
reason, we see an elongation of speckles due to their chromaticity even in an image nominally related to a single wavelength (see Fig.\ref{fig:psf}). 
We also note that the speckles seem to form a sort of bright ring: this is a
characteristic feature of XAO: the central part of the image is better
corrected than the external region, and the radius of this ``ring'' is $\sim
\lambda / 2d$ where $d$ is the interactuator distance of the wavefront
sensor. This radius is also called the Control Radius and is $\sim$0\farcs5 at
the shortest wavelength (0.95$\mu$m) in this application.


The PSF simulated in this way has been compared with real observed data 
obtained with NACO-SDI at VLT, adapting the input parameters for this specific
case. The results showed a good correspondence between the 
simulated speckle noise and the measured noise, and give confidence in the simulation \citep{ber}.

\subsection{The Integral Field Unit and the disperser}

The simulated IFU is a microlens array composed of 254$\times$254 hexagonal
lenses. For the 8 meter telescope, we used a spatial sampling of 0\farcs014 per
lens, covering a total field of view of 3\farcs5$\times$3\farcs5. To avoid stray light
from the gaps between the lenses, we applied a circular mask to every
lens. This mask has the disadvantage of blocking in the simulated case the 9.2\% of
light, but its presence is important to avoid stray light at the edges
of the lens. The simulated case is also the best case geometrically
possible: the radius of the circular mask $r_m$ is supposed to be equal to the
distance between the center of the hexagon and one of its sides. Mathematically
it means that $r_m = (l/4) \sqrt{3}$, where $l$ is the diagonal of the
hexagon. This loss of photons has some consequences on the procedure of
speckle subtraction, that have to be studied in detail, even if it is
considered in the code.

Each microlens of the IFU array samples a small portion of the PSF, and creates
its own small PSF which is imaged onto the detector by the IFS
optics\footnote{After the microlens array the diffraction of the lenslets is
  dominant: for this reason, we don't have micropupils on the detector but
  diffraction limited PSFs}. The
location of these PSFs depends on the geometry chosen for the array and
on the size of the lenslets. In CHEOPS IFS the optical design has been
optimized to keep the cross talk between the lenses below 10$^{-4}$, in order
to be negligible.  

The code considers each lens separately, applying a standard PSF given by the
diffraction pattern of a uniformly illuminated circular aperture:

\begin{equation}
i(\theta)=i(0) \left[ \frac{2J_1 \big( v(\theta) \big)}{v(\theta)} \right]^2 ,
\end{equation}

\noindent where $\theta$ is the distance from the center, $J_1$ indicates the
Bessel function at 1$^{st}$ order and $v(\theta)=2\pi \theta / (\lambda F)$
with $F$ being the focal ratio (in our specific case we used $F$=8).


The next step is the simulation of the disperser. Since the images we
simulated up to now (stellar PSF and microlenses) are related to a single
resolution element of the spectrum, thus almost monochromatic, we need to
repeat the entire procedure in order to simulate all the different
wavelengths of the considered spectrum. Every modification in wavelength causes two effects on the speckles on the AO focal plane: a radial shift with respect to the stellar center
and a change of the size of both the diffraction figures and the
speckles. Moreover, the PSFs of the microlenses will also change.
In addition, the fluxes of the star and of the planets must change with wavelength according to their spectra. Final images
after the simulation of the IFU for all different wavelengths must then be
combined on the detector in order to obtain the observed spectra, shifting the
micro-PSF images slightly in a way proportional to their wavelength. 

\begin{figure}
\epsscale{1.0}
\plotone{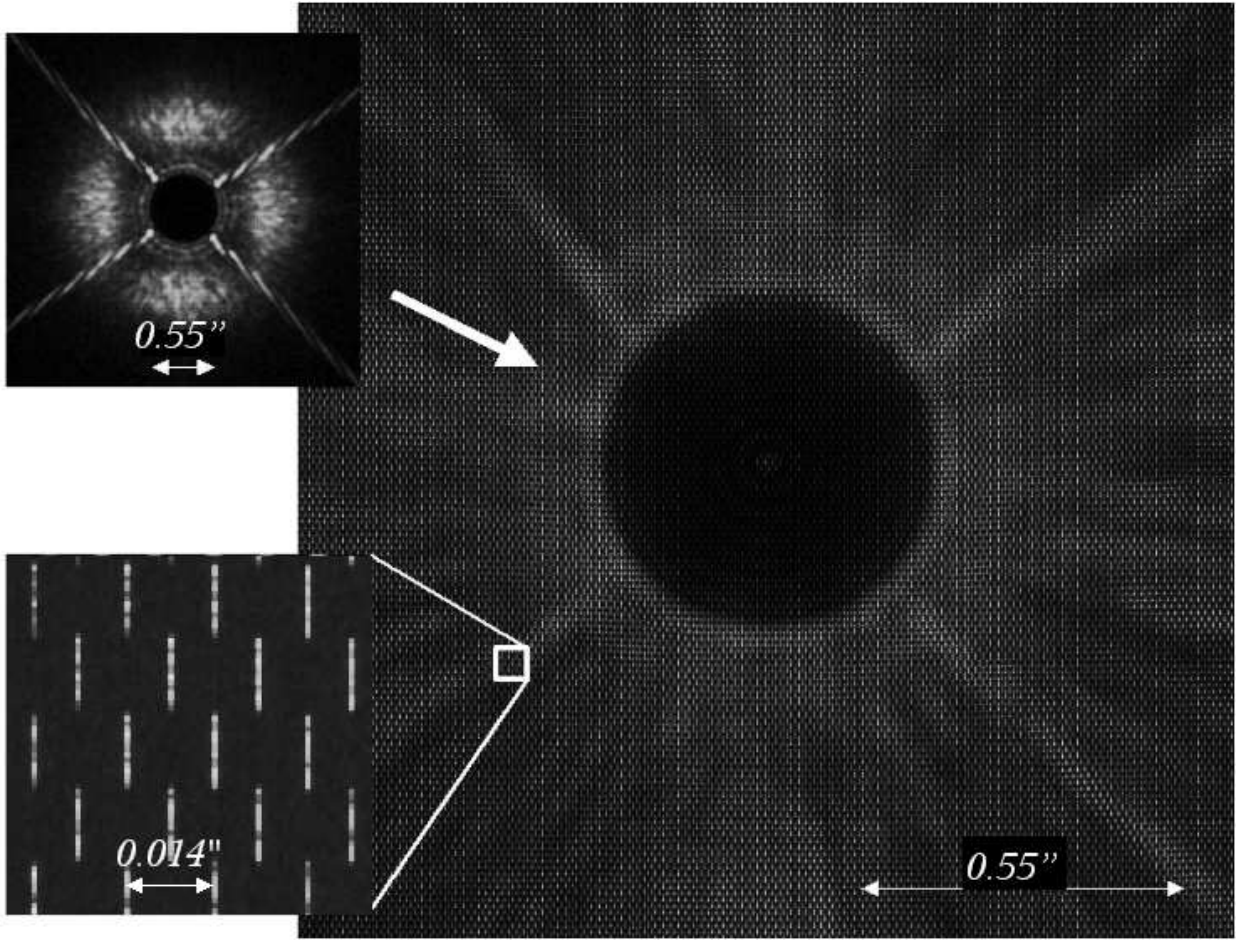}
\caption{Final output of an IFS observation of a star, in the case of the VLT and with the details described in the text. At top left, an example of PSF (with coronographic mask, spiders and AO correction) is presented. After the simulation of IFU and disperser we obtain the large image. At bottom left details of the spectra can be seen. Separations are expressed in arcesconds. The spectra cover the spectral region between 0.95 and 1.70$\mu$m, with a resolution of 15, so their spectral length is about 20 pixels. \label{fig:spettri}}
\end{figure}

Finally, before concluding the procedure, all noise sources related to the detector
are added to the image. These noise sources include read out noise, dark current, flat field etc. The contributions to
noise are computed considering the values appropriate for the main near-infrared
detector available at present. The used values for the various parameters characterizing the detector are summarized in Tab.\ref{tab:det}.
An example of the output of the whole simulation is presented in Fig.\ref{fig:spettri}.


\subsection{Data reduction}
\label{sec:reduction}

\begin{figure}
\epsscale{0.7}
\plotone{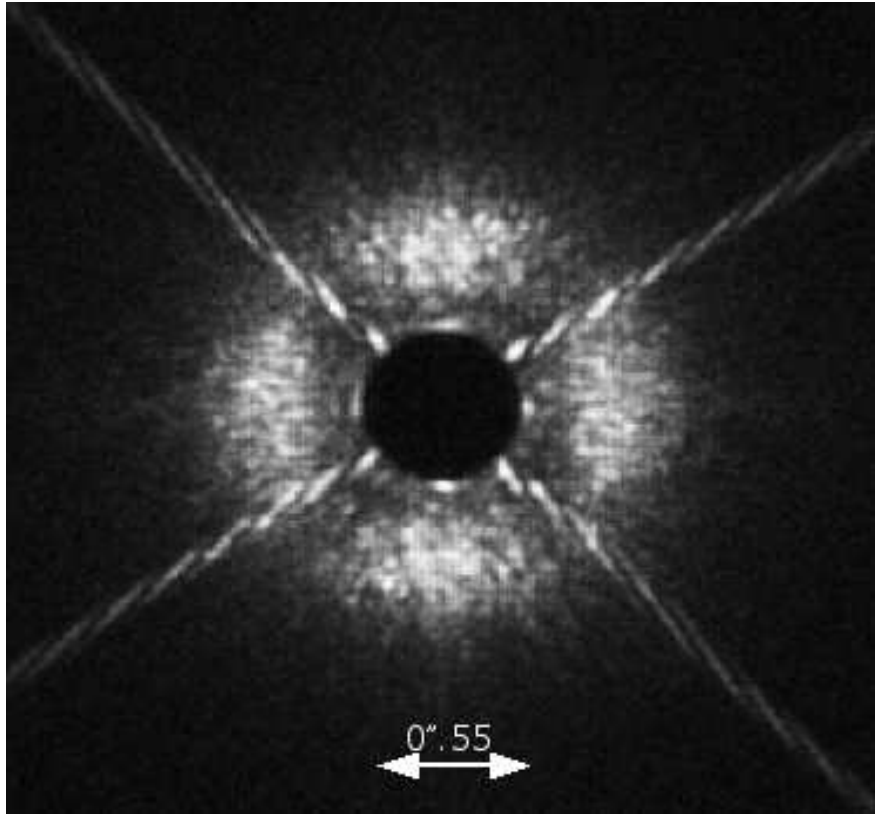}
\caption{A rebuilt monochromatic image obtained by the procedure of data
  reduction of the simulation code. As explained in the text, from the spectra
  we obtain a set of images like this that, after a scaling procedure in size,
  can be compared each other in order to enhance the planetary signal. \label{fig:rebuilt}}
\end{figure}

Fig.\ref{fig:spettri} clearly shows that the spectra in the final
image are not as regular as expected from the model of the stellar spectrum shown in
Fig.\ref{fig:star}. The dependence of fluxes on wavelength apparently follows a pseudo-sinusoidal law. This is due to the shift of speckles from
one microlens to another with changing wavelength. This effect makes it impossible
to do a simple comparison of the spectra for detecting the planet. For this reason,
following the example of \citet{spa}, we must extract all data related to every
single resolution element from all the spectra in order to rebuild a set of
monochromatic images that can be, at this point, rescaled in size by a factor
equal to $\lambda_0 / \lambda$. Scaling all the monochromatic images at a specific
$\lambda_0$ makes the speckles coincide in position and size, so that a
subtraction becomes possible. An example of rebuilt monochromatic
image is shown in Fig.\ref{fig:rebuilt}.


The algorithm for subtraction can be different depending on the nature of the planet. In this first analysis, we decided to follow the idea of double-difference presented by \citet{mar}, using three images $I_1$, $I_2$ and $I_3$ obtained
by three resolution elements $\lambda_1$, $\lambda_2$ and $\lambda_3$:

\begin{equation}
\label{eq:dd}
D = \left( I_1 - I_2 \right) -k \left( I_1 - I_3 \right),
\end{equation}

\noindent where $k$ is a constant factor depending on the Strehl ratio achievable at the different wavelengths considered:

\begin{equation}
k= \frac{S_1 - S_2}{S_1 - S_3}.
\end{equation}

\noindent 
With a number of wavelengths available larger than 3, we could apply
Eq.\ref{eq:dd} to various combinations of images, investigating different spectral features and using the entire
information given by the spectra, as explained in Sec.\ref{sec:why}. 

\begin{figure}
\epsscale{1.0}
\plottwo{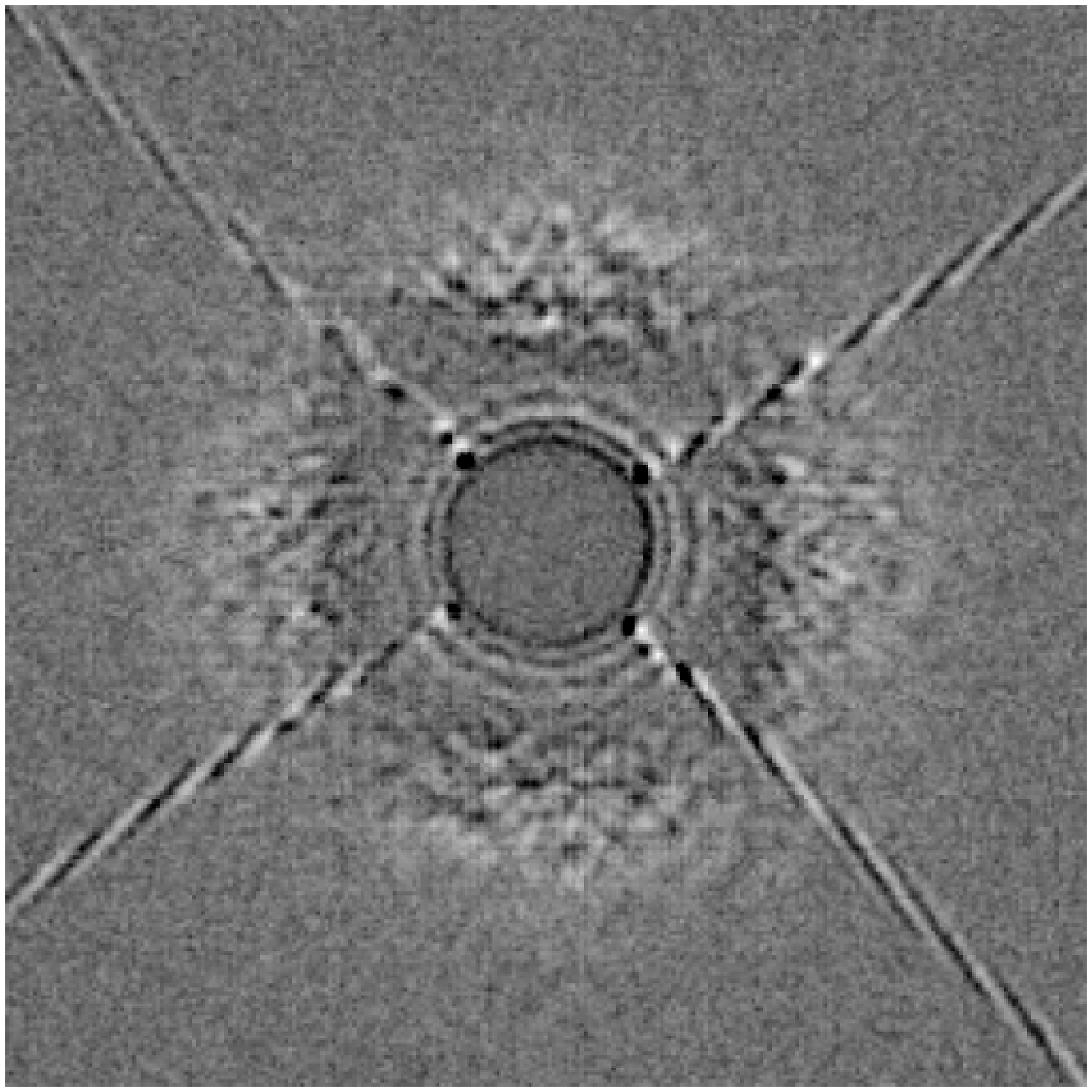}{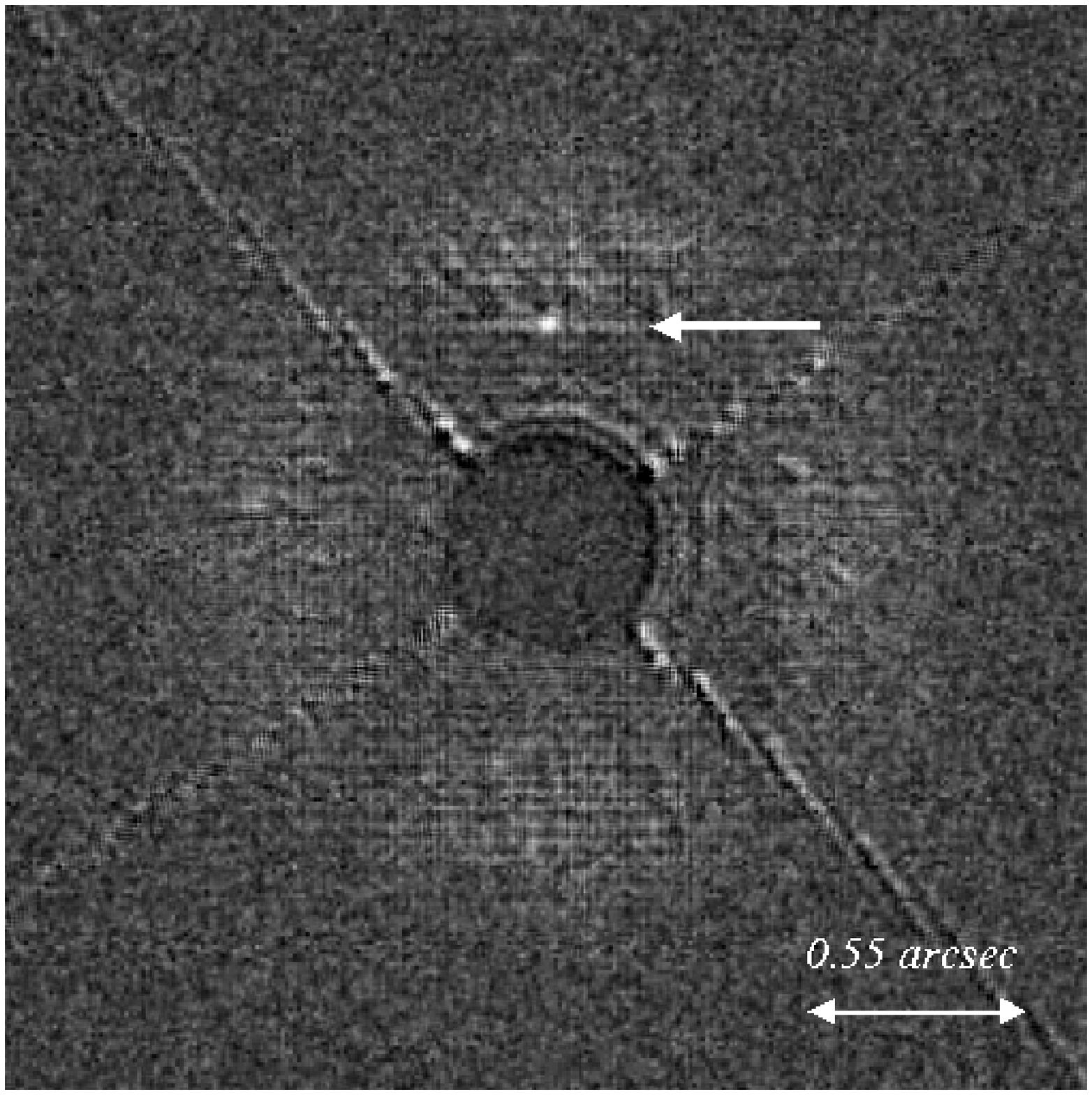}
\caption{Detection of a simulated brown dwarf of 30\,M$_J$ after 1 hour of exposure time, around a 5\,Gyr old G0V star at 40\,pc. Left: three images obtained at different wavelength, on and off a methane absorption band, are combined using Eq.\ref{eq:dd} but without an accurate spatial scaling: the brown dwarf is visible, but at a signal-to-noise ratio low enough to confuse it with a false alarm. Right: the results of the same operation, but now rescaling appropriately the images by a factor $\lambda_0 / \lambda$ (see Sec.\ref{sec:reduction}). An appropriate scale factor has been applied also to the fluxes. The result is a clear unambiguous detection of the companion, not possible without application of the rescaling procedure, with a signal-to-noise ratio of $\sim$30. \label{fig:bd}}
\end{figure}

An example of detection of a brown dwarf of 30\,M$_J$, 5\,Gyr old, orbiting around a solar-type star, is shown in Fig.\ref{fig:bd}. As mentioned above, the mechanism for reduction of speckle noise presented here is useful only if
the monochromatic images obtained from the spectra are scaled in an appropriate
way: no rescaling (at left) usually makes
the companion not detectable, while with a good rescaling procedure (at right) the
companion is clearly visible with quite a high signal-to-noise ratio ($\sim$30).
 
\subsection{Simulation code -- Next steps}

The CSP code, although it takes into account a large number of parameters and
noise sources in order to simulate real observations, can
still be improved. Some particulars have been neglected or underestimated
for various reasons, sometimes because going too much into details was not the
purpose of the work presented here (e.g. the AO system structure), and
sometimes because 
we were interested in immediate performances estimates while a more accurate
study would have required excessive computer time. This study will be anyway
investigated in detail later.

In particular, the procedure for simulating the microlens array presents some simplifications, and a possible improvement is under study now. At present, this
procedure supposes all the lenses to be identical and it associates a standard perfect PSF to each lens, but in reality the phase screen over the entrance pupil of a
microlens is not constant. A first set of tests and simulations showed that
with our specific choices of focal ratios and spatial sampling, for low static
aberrations, the differences between the real observed PSFs of the microlenses
and the simulated perfect ones can be neglected for long exposure times, when
the speckles are homogenized to form a smooth seeing disk. A study of these differences in the case
of significant static aberrations is ongoing.

Another phenomenon that is currently not well accounted for, is
the interference between the microlenses. This interference could modify the
shape of the PSFs, making them sometimes too faint to obtain a clear spectrum. An accurate simulation of this effect is very expensive in terms of computer resources, and had to be neglected in this first version
of the code. 
However this is currently being studied in order to determine whether omitting this effect is significant when achieving accurate detections. The results
of these studies will be published in the near future.

To be included in our simulations, and consequently in the procedure of
analysis of spectra, is a calibration error due to the fact that in the real
observations the central wavelength related to every spectral pixels of each
spectrum cannot be perfectly known but requires a calibration; in our code
this error is probably underestimated, and new studies are ongoing now to understand its importance in the final signal-to-noise ratio of the detections.

Concerning the flat field, one should note that the whole process of generating noise is not simulated in detail. We simply calculated a residual noise of 10$^{-4}$, as explained
in Sec.\ref{sec:crosstalk}. A complete analysis of the flat field process could
be important for understanding their effect on the final signal-to-noise ratio
achievable. 

Finally, during or between the real observations we will need to apply some rotations to
the field of view. An estimate of the level of noise introduced by these rotations and the following corrections could well be interesting. Moreover, anisoplanatism
can also introduce some reductions to the Strehl Ratio of the planet that
should be not very significant but have to be studied in detail.

\section{Results}
\label{sec:results}

\subsection{Compilation of results}


Here we present the results of the simulations we made in the case of an
observation of various Jupiter-like planets and low-mass
brown dwarfs with an 8m-class telescope. For each case we calculated a signal-to-noise ratio, in order to
define which kinds of planets are detectable with a signal-to-noise ratio larger than 5, the threshold of detectability. We simulated the following specific cases:
\begin{itemize}
\item 4 different planetary masses: 1, 5, 15, 30\,M$_J$.
\item 3 values of ages: 0.01, 1, 5\,Gyr.
\item 3 distances from the Sun: 3, 10, 40\,pc.
\item 2 types of stars: a G0V and a M0V.
\item 3 different seeing conditions: 0.60 (good), 0.85 (median), 1.00 (not so
good) arcseconds.
\item Various angular separations between the host star and the companion, over a
range between $\sim$0.3 and $\sim$2 arcseconds.
\end{itemize}
The exposure time was always assumed to be 4 hours.

In this list of cases we found objects warm enough to be characterized only by
intrinsic thermal emission, and at the same time old and small planets that
shine only due to reflected light. Moreover, usually the planetary spectra are different
as a function of the physical separation between planet and star. Thus,
the sample we covered is appropriate for a wide group of targets within the 
capabilities of an 8-meter telescope. We needed a large set of spectral models, that were taken from \cite{bur}.

The main results of the simulations are presented in this section. As secondary results, we also discuss the general capabilities of this technique in
removing speckle noise (Sec.\ref{sec:epsilon}), and the frequency
of false alarms (Sec.\ref{sec:false}).  

In Tab.\ref{tab:plotg} we compiled the signal-to-noise ratios for all the cases listed above, for good seeing conditions (seeing of 0\farcs6).
The values are expressed as a function of the angular separation from the host
star.
It is important to note that we tried to be conservative in
choosing the various parameters, in order to present a lower limit of the
expected signal-to-noise ratios.

\begin{figure}
\epsscale{0.9}
\plottwo{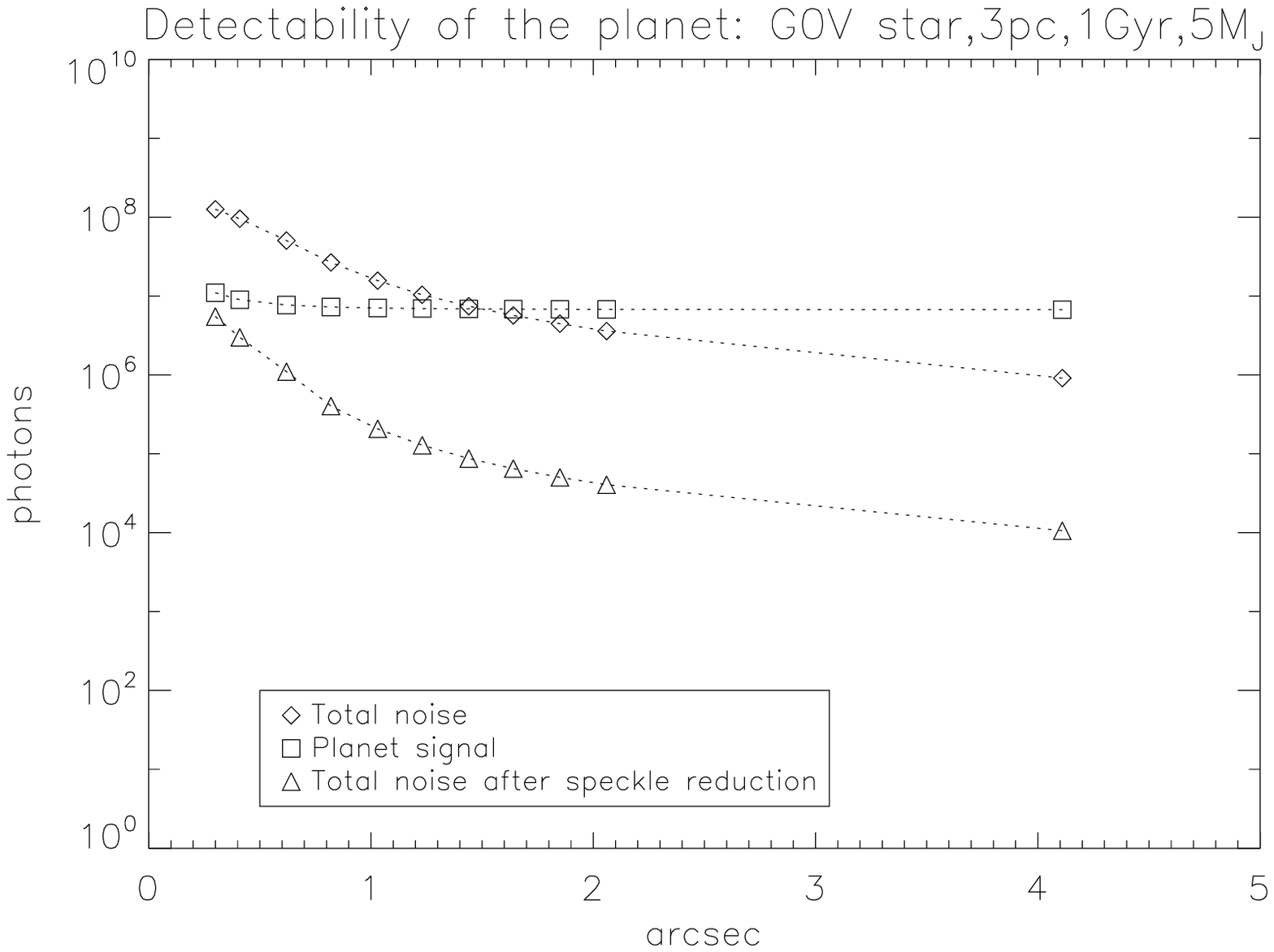}{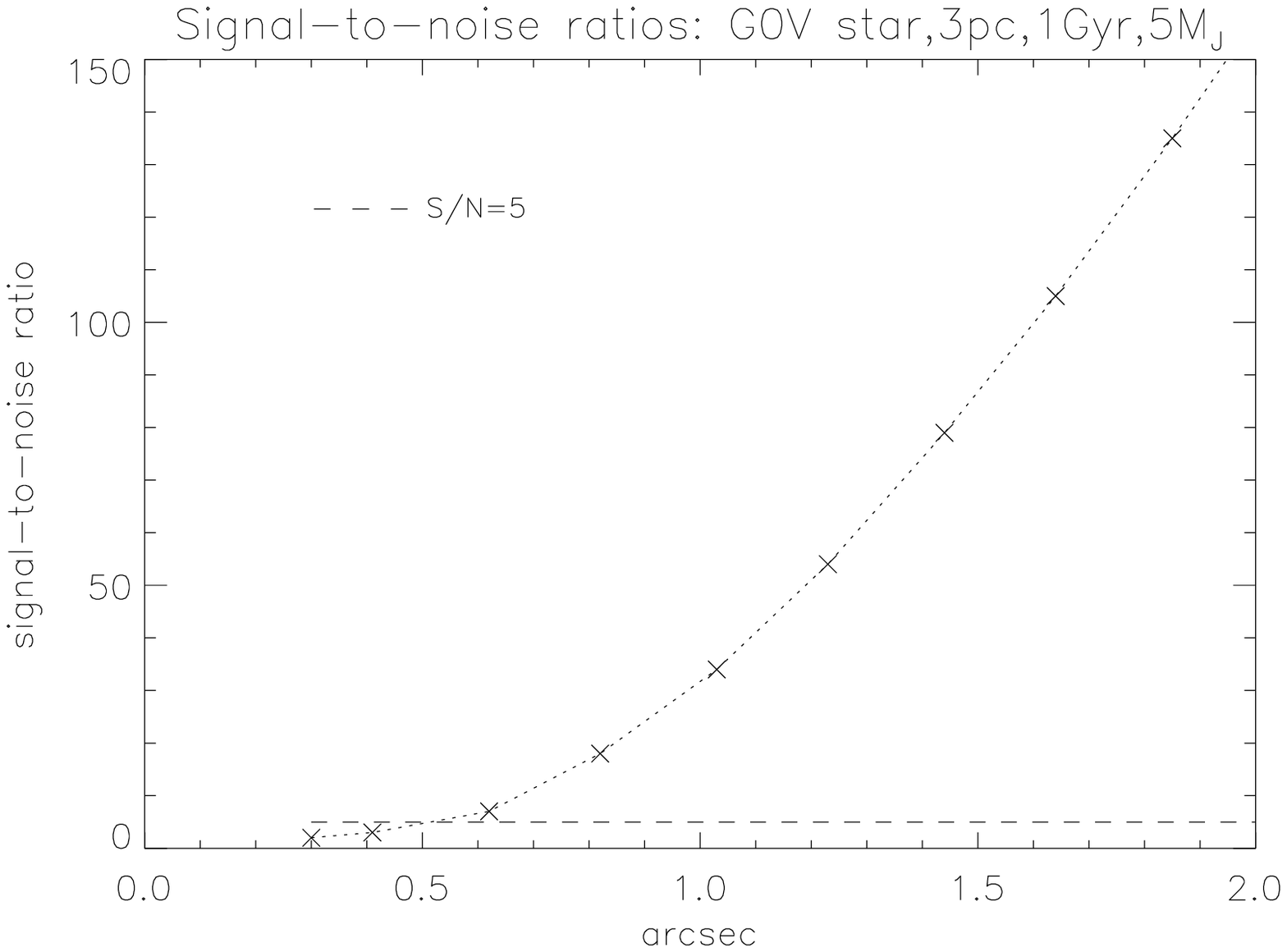}
\plottwo{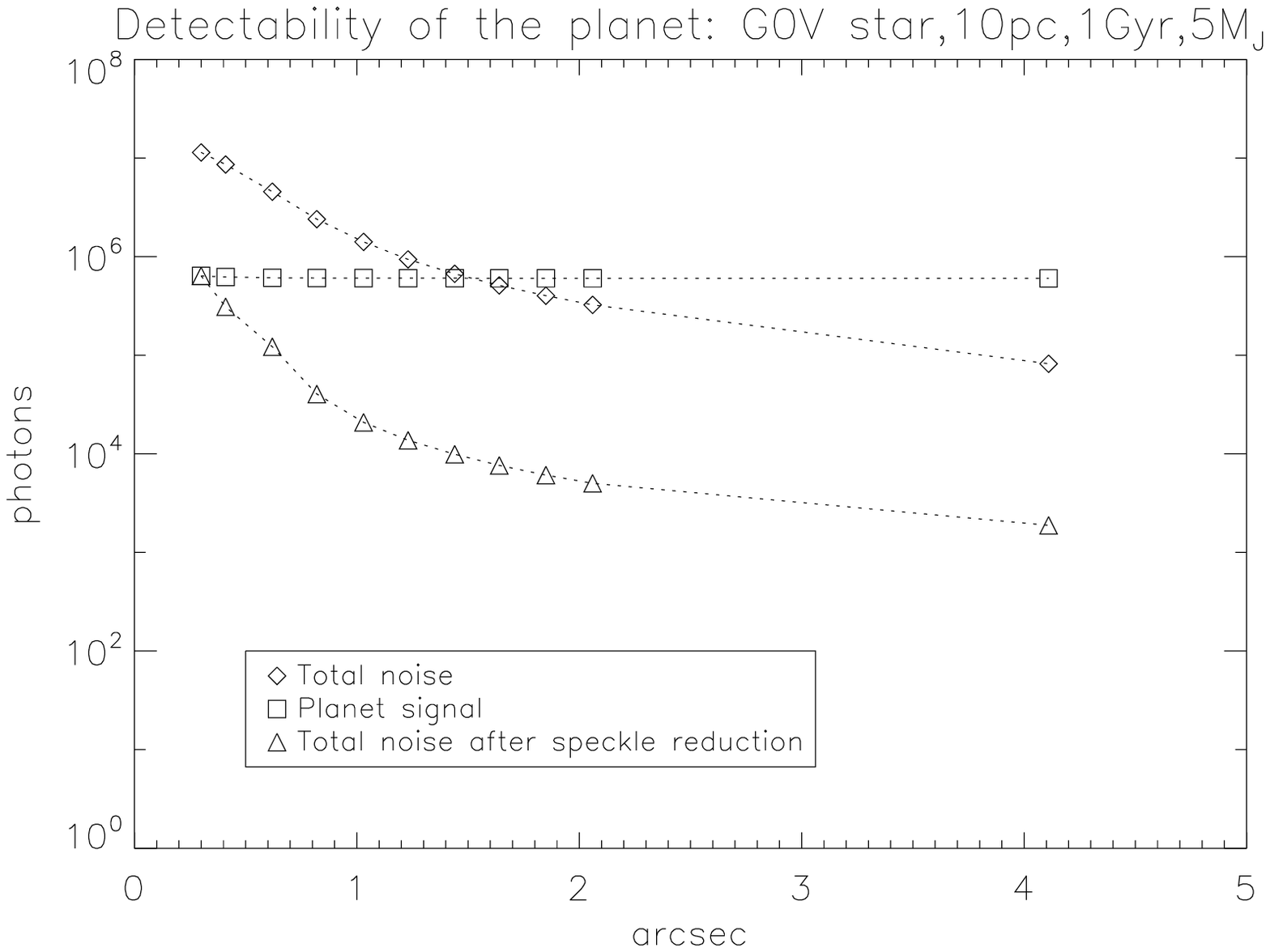}{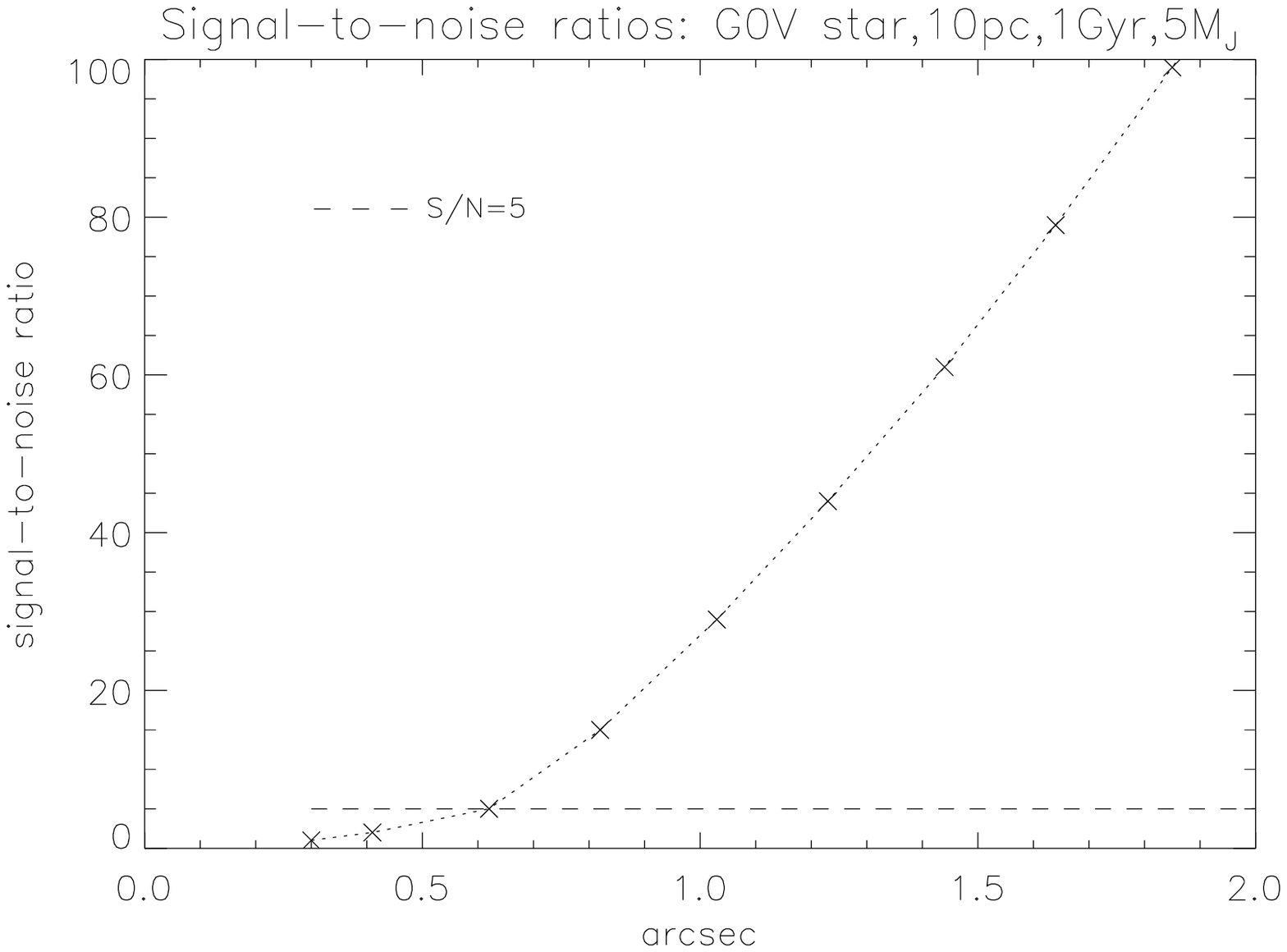}
\plottwo{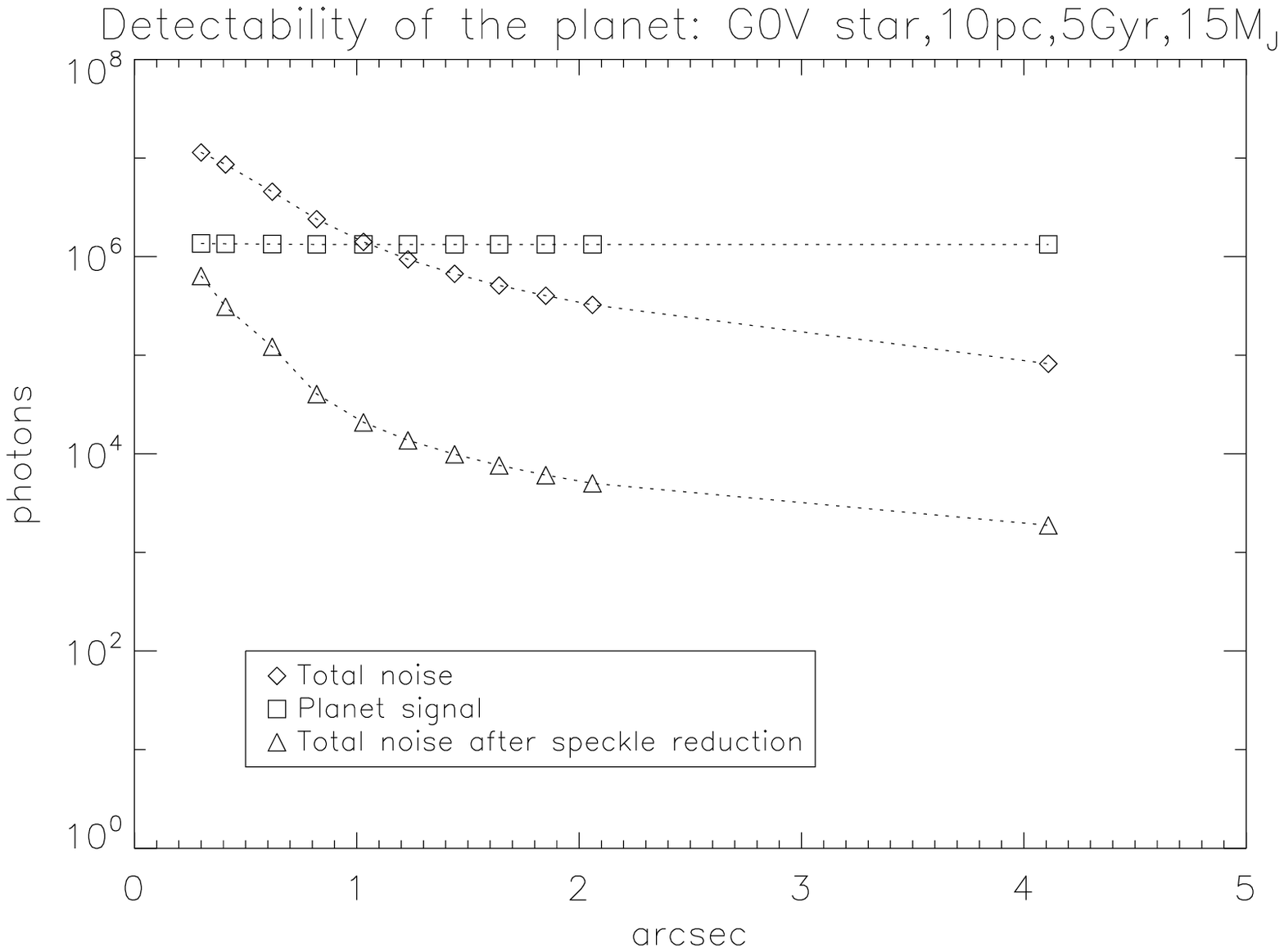}{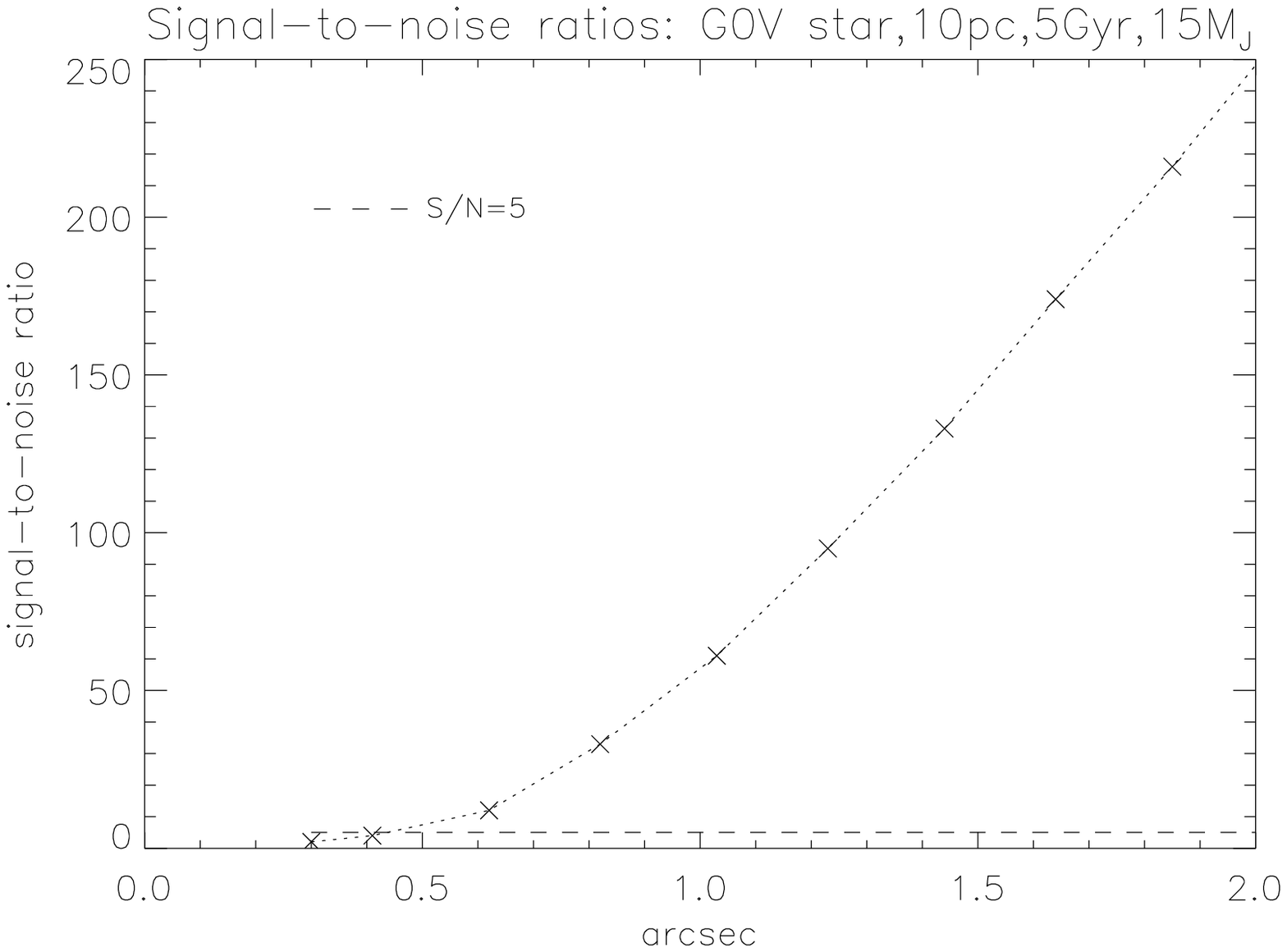}
\plottwo{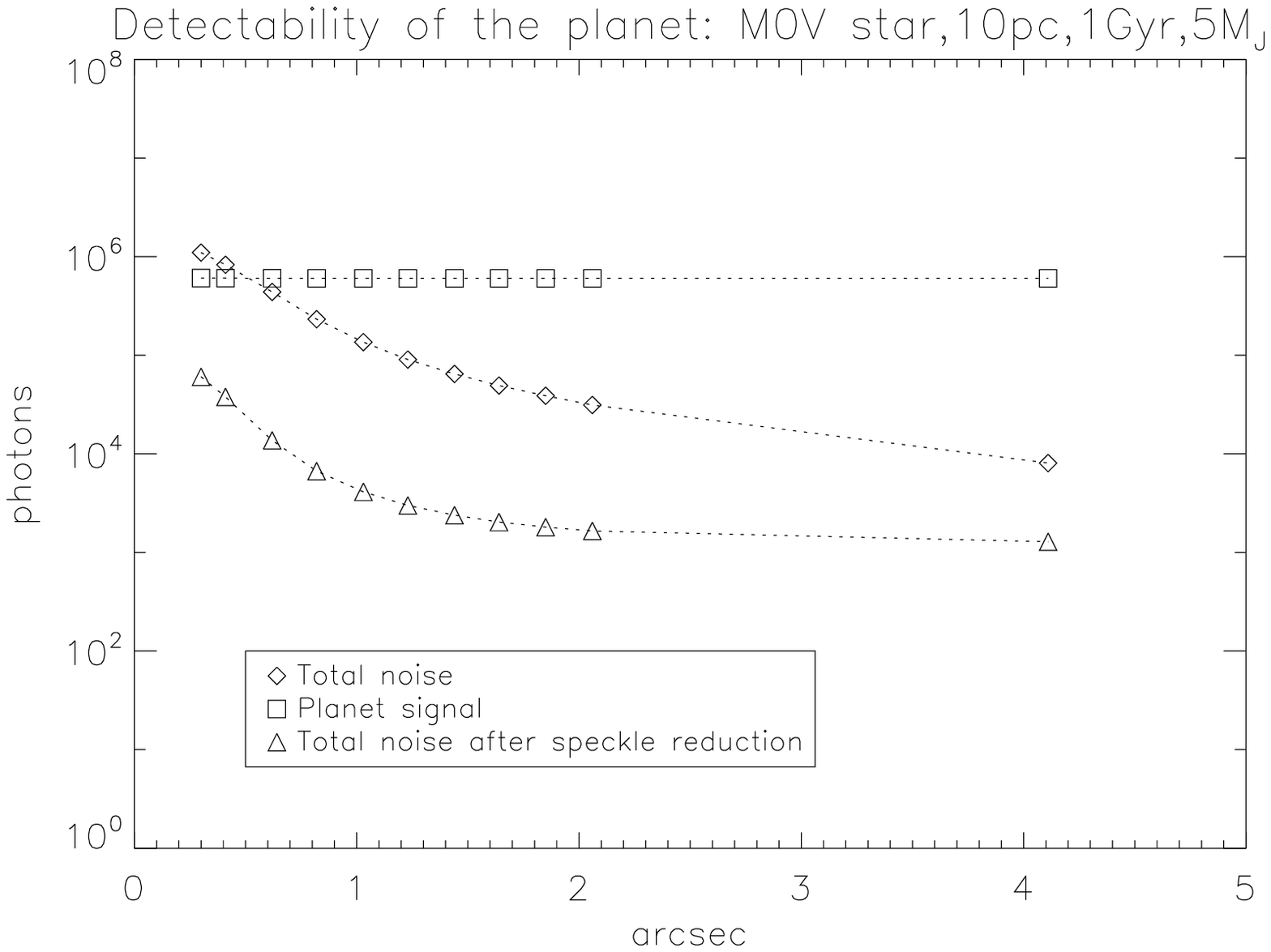}{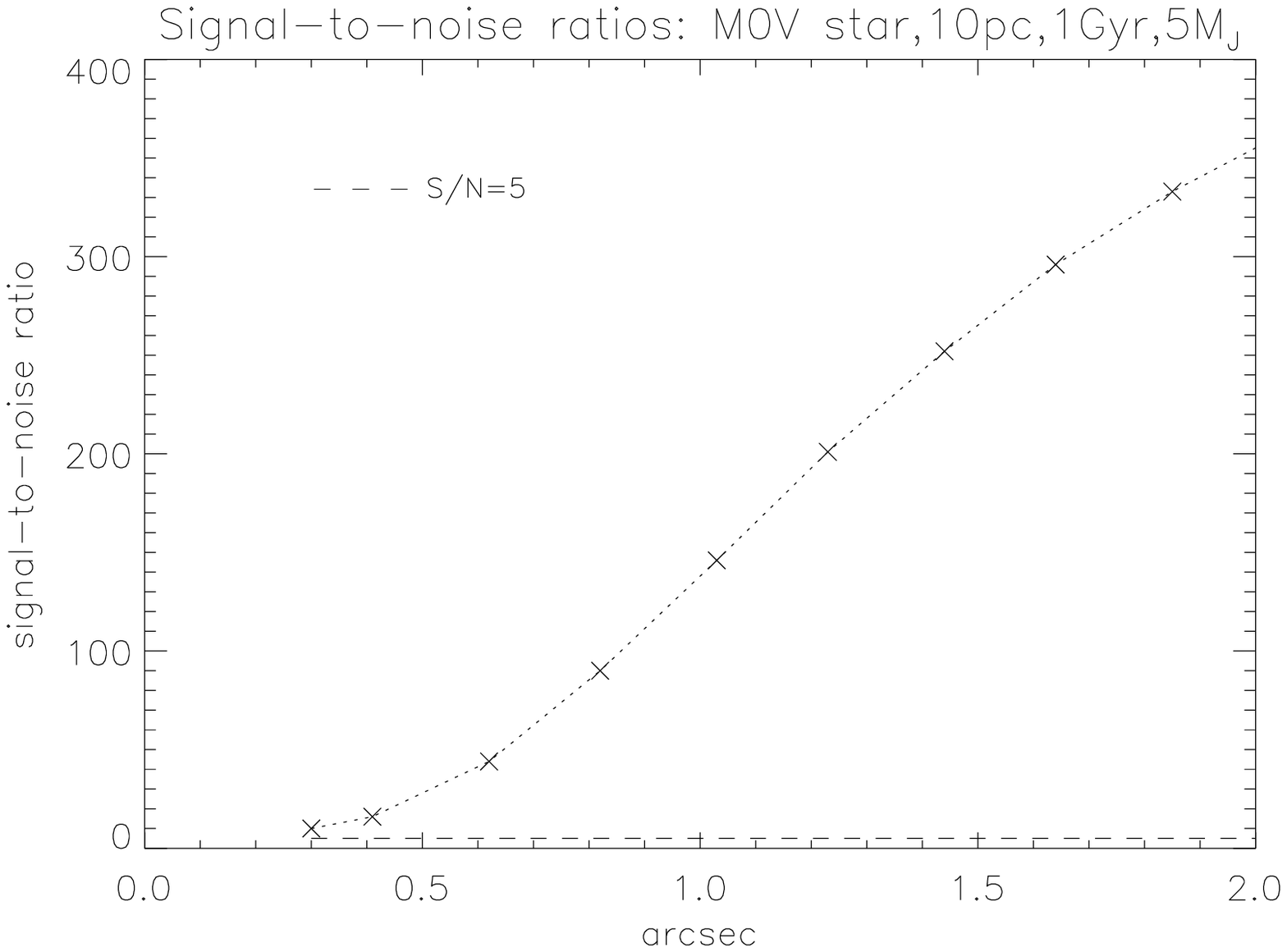}
\caption{Left: The levels of signal and noise are shown, before and after SDI, for four cases. Right: The derived signal-to-noise ratios. The cases are, from top to bottom: G0V star, 3\,pc, 1\,Gyr, 5\,M$_J$; G0V star, 10\,pc, 1\,Gyr, 5\,M$_J$; G0V star, 10\,pc, 5\,Gyr, 15\,M$_J$; M0V star, 10\,pc, 1\,Gyr, 5\,M$_J$. \label{fig:snplots}}
\end{figure}

For some significant cases we also show plots (Fig.\ref{fig:snplots}, left) with
the level of noise before and after the subtraction
of speckles, the photon noise and the expected signal from the planet. Plots
of the measured signal-to-noise ratios versus separations are also shown (Fig.\ref{fig:snplots}, right).


Because of the expensive calculations to simulate observations of 4
hours with our code, some of these results have been
obtained using scaling laws and a faster spreadsheet which computes the signal-to-noise of a planet
image at different separations from the star as expected at the output of a 
simultaneous differential imager like the one we simulated. This spreadsheet,
called CSS (CHEOPS Simulation Spreadsheet) is based both on theoretical estimates and on the simulated PSFs and noises. Many cases presented in Tab.\ref{tab:plotg} have been obtained using both the simulation code and the spreadsheet: the good coincidence of the results
allowed us to use it for the detections not easy to simulate: for
instance, the objects at angular separation of 4\farcs21 which cannot be simulated because it is out of our field of view. A description of the CSS spreadsheet is presented in
Appendix \ref{app:css}.

\subsection{Determination of the factor of reduction of speckle noise}
\label{sec:epsilon}

A crucial result of these simulations is the determination of a value 
characterizing the effective reduction of speckle noise achieved using the SDI
technique as suggested in this paper. We know from theory the behaviour of
speckle noise as a function of time, flux and Strehl ratio (Eq.\ref{eq:var}).
Applying SDI to the obtained data reduces the noise by a certain
factor which we call $\varepsilon$. The expression for the reduced speckle noise $N$ would be
this:

\begin{equation}
N = N_{sp}\varepsilon = \sqrt{1.7 \pi \frac{\tau_0}{t}} \left[ \left( 
\frac{\lambda}{D} \right)^2
\left( 1-S \right) F_* f_{halo}(\theta) \right] \varepsilon(\theta),
\end{equation}

\noindent where $N_{sp}$ is the original
level of speckle noise and the other variables are as explained in Sec.\ref{sec:speckle}.  In order to calculate $\varepsilon$, we considered first 
one of the monochromatic images, before the SDI procedure, scaled to the 
wavelength $\lambda_0$ of the original image. In this image we analyzed the
level of noise $N_{sp}$ in a set of small regions placed in a radial sector, 
at different angular separation from the central peak. This radial sector has been
chosen to avoid spiders, artifacts and also the simulated planets.
 We repeated the procedure using the
image resulting from the SDI, obtaining a resulting noise $N_d$.
While in $N_{sp}$ the speckle noise is strongly dominating, in $N_d$ the
contributions of photon noise $N_{ph}$ and of other background noise sources ($B$, 
that are detector noise, background sky, etc.) become important. These contributions can be
easily derived from the average intensity in the regions analyzed before
subtracting for $N_{ph}$, and by measuring the level of noise in a peripherical
region not affected by speckles. 

The formula we used to calculate $\varepsilon$ is finally, for each portion:

\begin{equation}
\varepsilon (\theta)= \sqrt{ \frac{N_d^2(\theta) - N_{ph}^2(\theta) - B^2}
{N_{sp}^2(\theta)}}.
\end{equation}

\noindent The quantity $\theta$ indicates the distance between the center and the different region of the image. Since the
regions are chosen on a radial sector at different separations from the
center, $\theta$ is expressed in arcseconds. We repeated the procedure for different sizes of the regions and for 
different radial sectors. The results obtained are sometimes quite different
because the simulated PSF is not perfectly symmetric. They were averaged in
a final $\varepsilon (\theta)$. 
 
\begin{figure}
\epsscale{.50}
\plotone{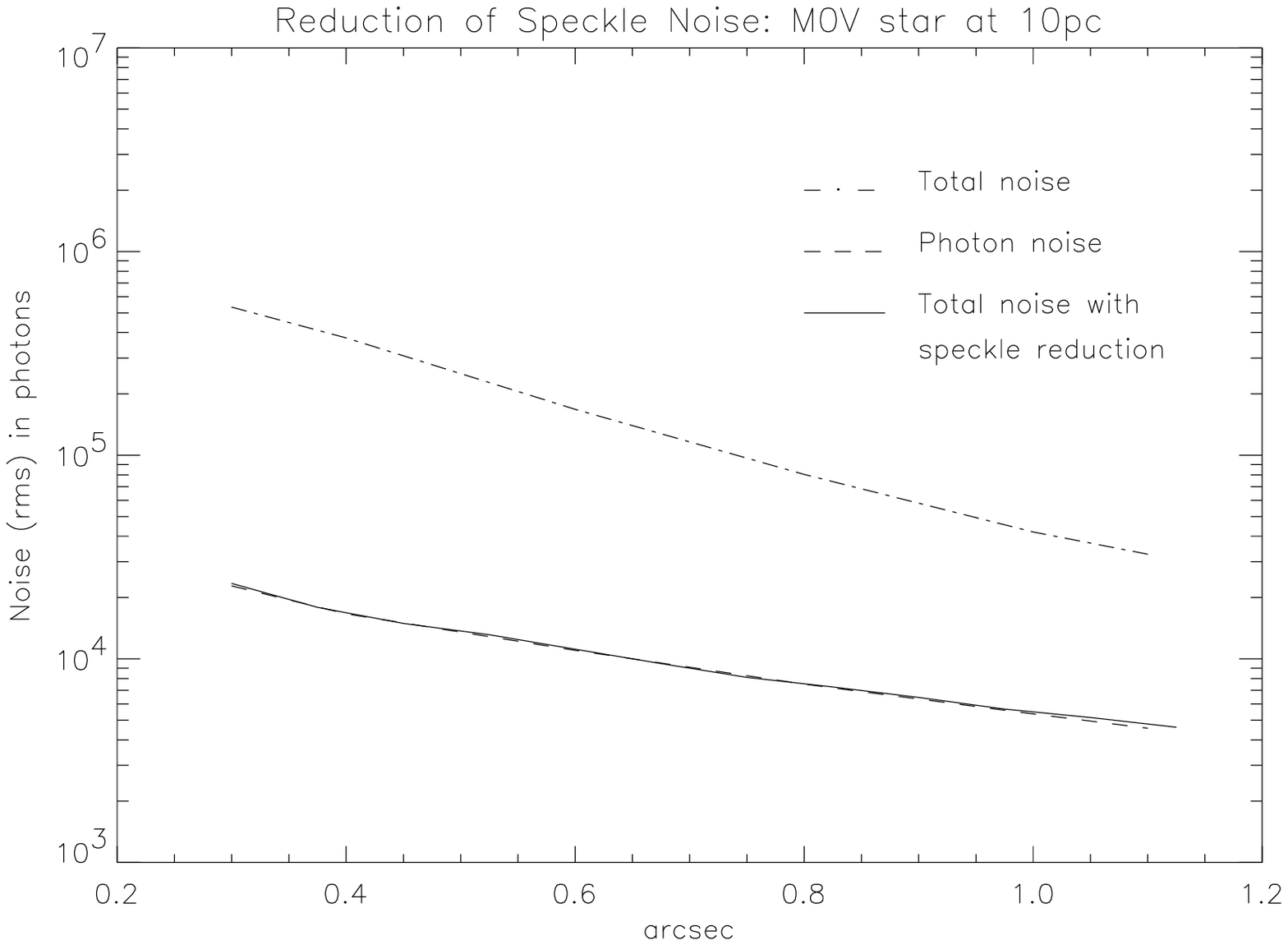}
\plotone{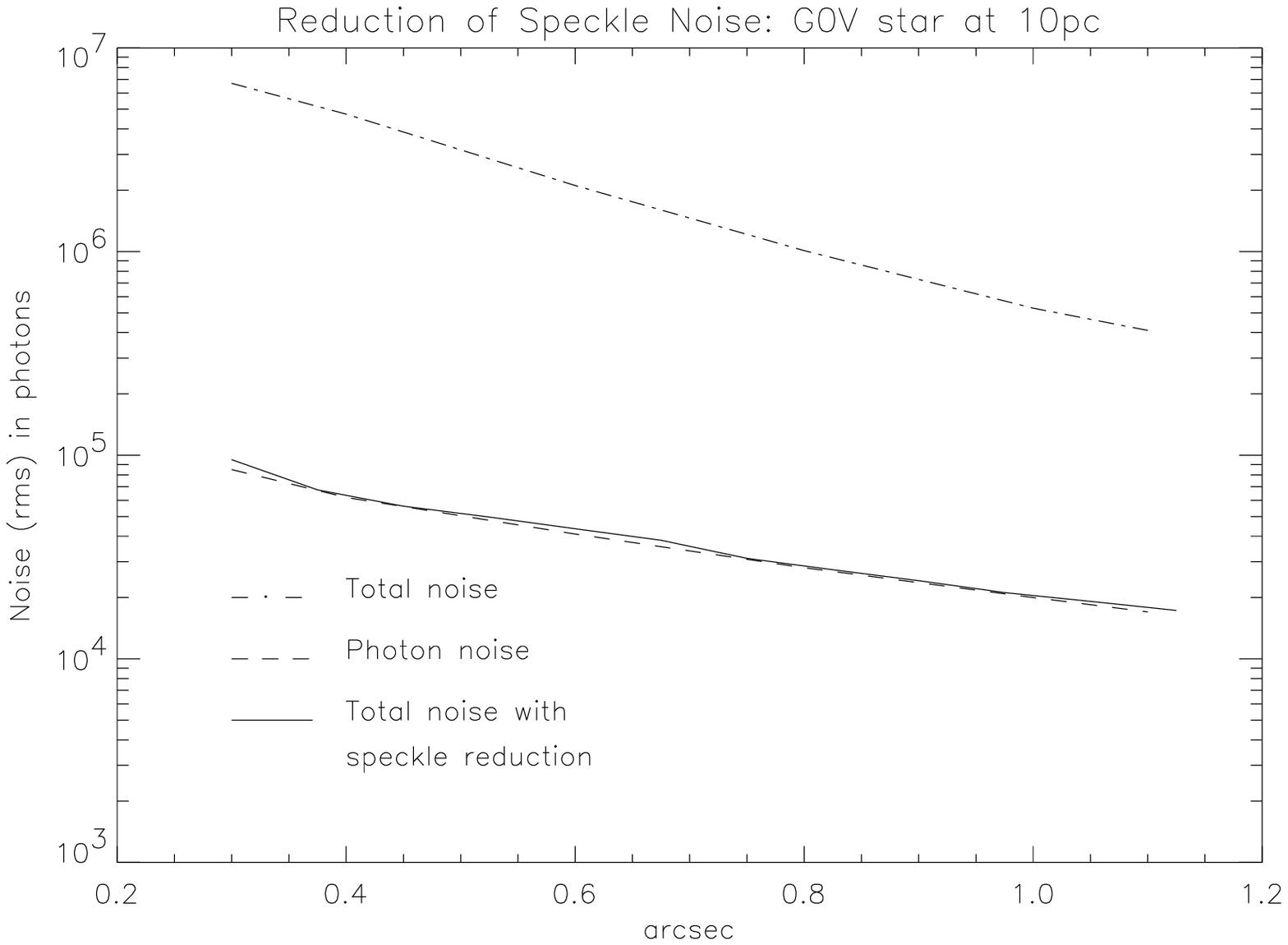}
\plotone{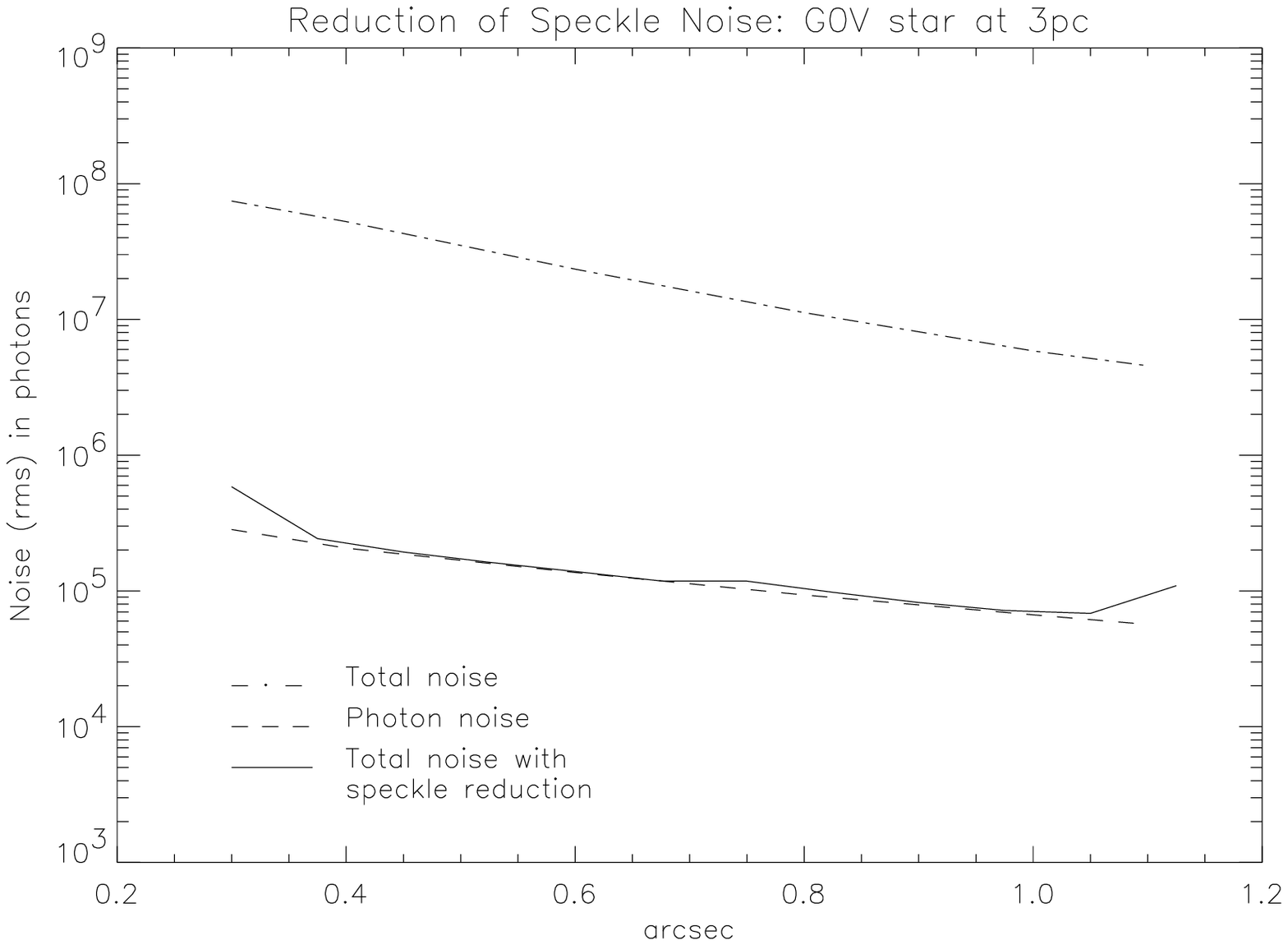}
\caption{Reduction of the speckle noise as a function of the angular separation from the star: Dash dotted line represents the total level of noise before SDI, dominated by speckle noise. The dashed line is the photon noise. The solid line is the total noise after application of SDI. Top: the case of an M0V star at 10\,pc from the Sun. Center: a G0V star at 10\,pc. Bottom: a G0V star at 3\,pc. In all cases we obtained reductions of a factor around 10$^{-3}$, and the final level of noise is very close to the photon noise. \label{fig:spkM0V}}
\end{figure}

From Fig.\ref{fig:spkM0V} 
we can easily see
that the final noise using this
technique becomes comparable with the photon noise for the three cases considered (M0V star at 10\,pc, G0V star at 10\,pc, G0V star at 3\,pc).
The factor of reduction changes with the separation, but it is
$\sim$10$^{-3}$ between 0\farcs3 and 1\farcs0. This value for $\varepsilon$ is
an indicator of a good suppression of speckle noise. 

This result shows that using the SDI technique is actually possible to reach a
level of noise comparable with the photon noise. Similar conclusions, in terms
of dual band imaging SDI, have been achieved for NACO-SDI in \citet{bil1}.


\subsection{False alarm frequency}
\label{sec:false}

We estimated the probability of detecting a false signal with a signal-to-noise 
ratio between 1 and 7, with a sampling aperture
corresponding to $\sim$2$\times$2 spatial resolution elements. Using this 
probability, we could extrapolate the expected number of events in the final differential image. The results are presented in Tab.\ref{tab:false}. These probabilities clearly demonstrate that the
threshold signal-to-noise of 5 is high enough to avoid false alarms. 
Therefore, we choose this value in an evaluation of detecting planets with the
IFS.
Moreover,
a more confident confirmation of detections can be done repeating the
observations under the same conditions if possible, and at different
angles. Comparing the position of the detected candidates, we can easily
confirm or not if it is a false alarm.


In the case of a large number of observations the number of false alarms can
become more important. 
With a S/N of 5, we expected one false alarm within 30 observations. This means that a ``detection'' always has to be checked by additional observations.


\subsection{Flat field errors and thermal background}
\label{sec:crosstalk}

\begin{figure}
\epsscale{.70}
\plotone{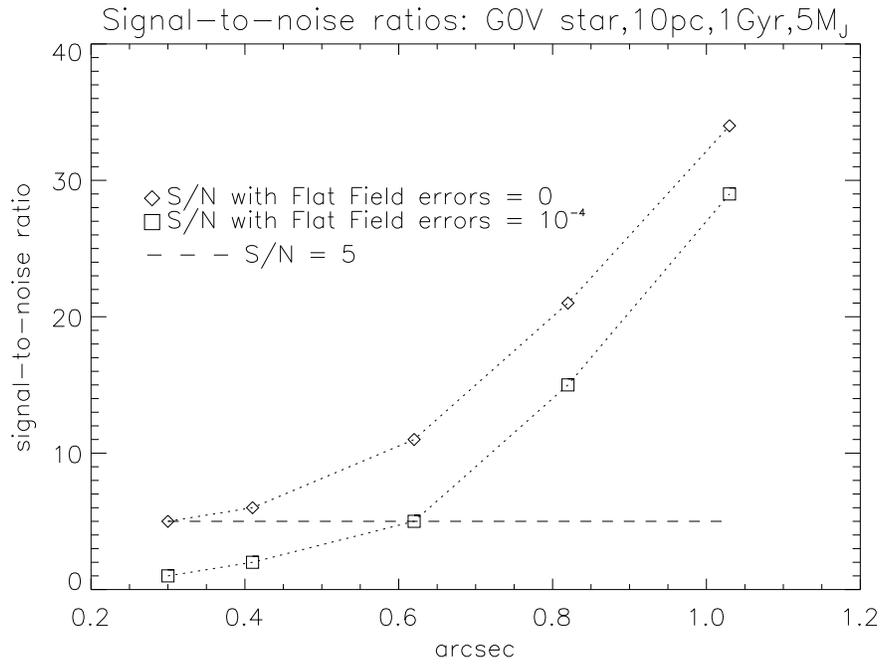}
\caption{Importance of flat field errors in detecting companions. In the case of a 15\,M$_J$ object, around a G0V star (age 1\,Gyr and at 10\,pc from the Sun), the signal-to-noise ratios decrease below 5 (dashed line) in the region closest to the star.
\label{fig:ff}}
\end{figure}

Flat field errors were included as multiplicative terms to the total number
of photons detected by each pixel of the detector. Accurate flat fielding is foreseen and images will be obtained using
a dithering procedure that spread the signal of the planet over a large number
of pixels by allowing the field to rotate from one exposure to the following
one (typically, a few thousands of exposures are summed up to provide the total
exposure). Therefore it is reasonable to assume that the residual flat field
error in a detector pixel can vary between
$10^{-3}$ that can be achieved using an accurate dithering procedure, and the
goal of $10^{-4}$ . This value is assumed in our computations,
although it is certainly not easy to achieve. Flat field errors are important for bright sources and for the regions
closer to the stars, limiting sometimes the possibility of observing 
planets by reflected light when using the SDI-IFS. Fig.\ref{fig:ff} shows this
comparison in terms of measured signal-to-noise ratios for a 15\,M$_J$ object
orbiting a G0V star at 1\,Gyr age at 10\,pc from the Sun: the detections within 0.3 and 0.6 arcseconds, which was possible without considering flat field errors, are shifted below the threshold of S/N=5.

Concerning the thermal background, which is also a source of errors, we 
assume that cold edge filters are placed in front of the IFS detectors. These edge filters will reduce the
thermal background flux by a factor between $10^{-4}$ down to
$10^{-5}$. We adopted this last value in our computations. The
thermal background flux is not a serious concern in the $J-$band but it may be
of some importance for the faintest sources in the $H-$band.

\subsection{Dependence of detection quality on seeing conditions}
\label{sec:seeing}

To show the effect of median (0\farcs85) or relatively bad (1\farcs00) seeing conditions in comparison with the good case described above, we
repeated the simulations for a few specific planets, selected in order to avoid
too
bright or too faint planets. From these simulations it becomes obvious how important it
is to work with the best atmospheric conditions when using our method to detect planets.

\begin{figure}
\epsscale{1.0}
\plottwo{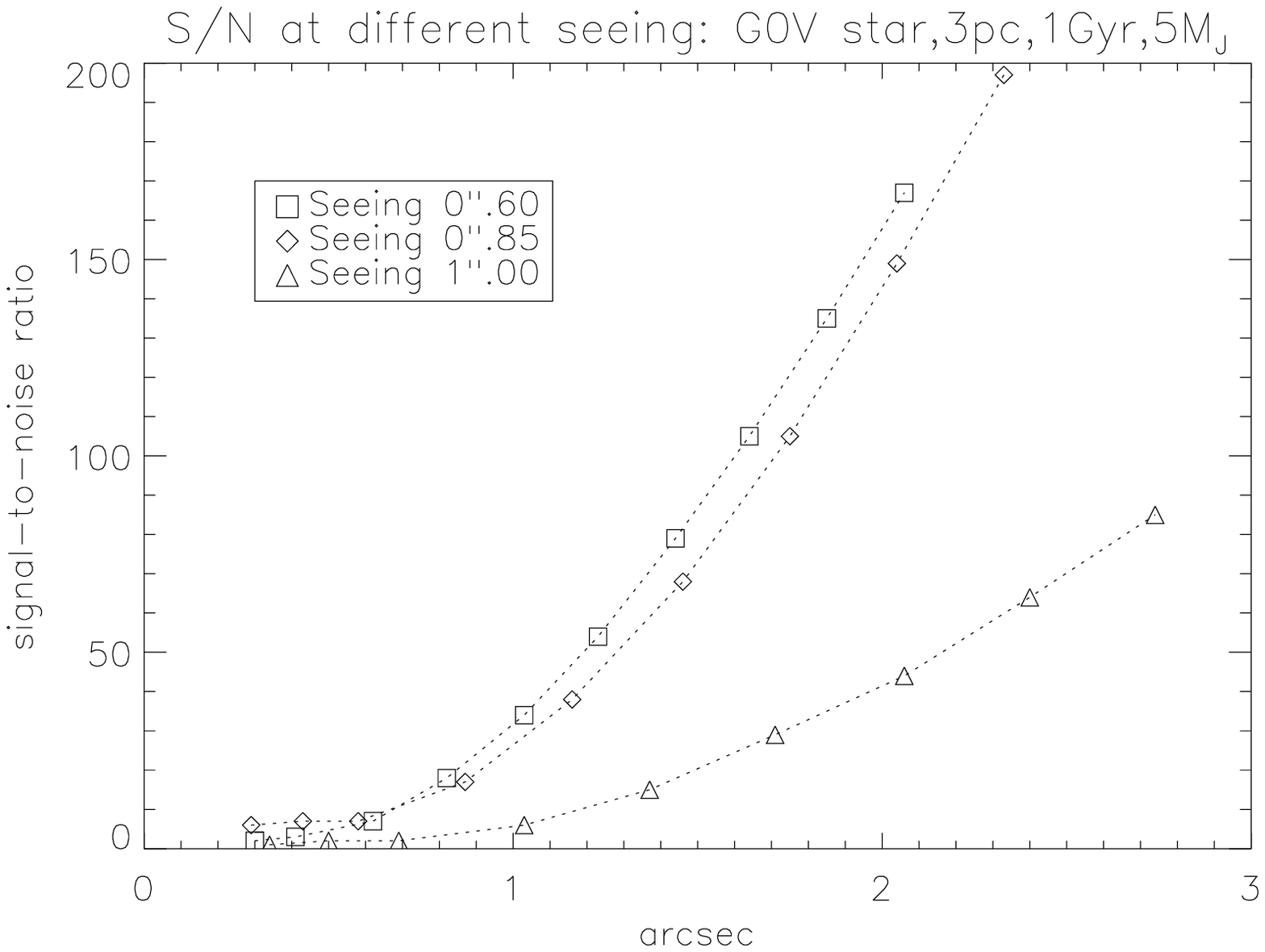}{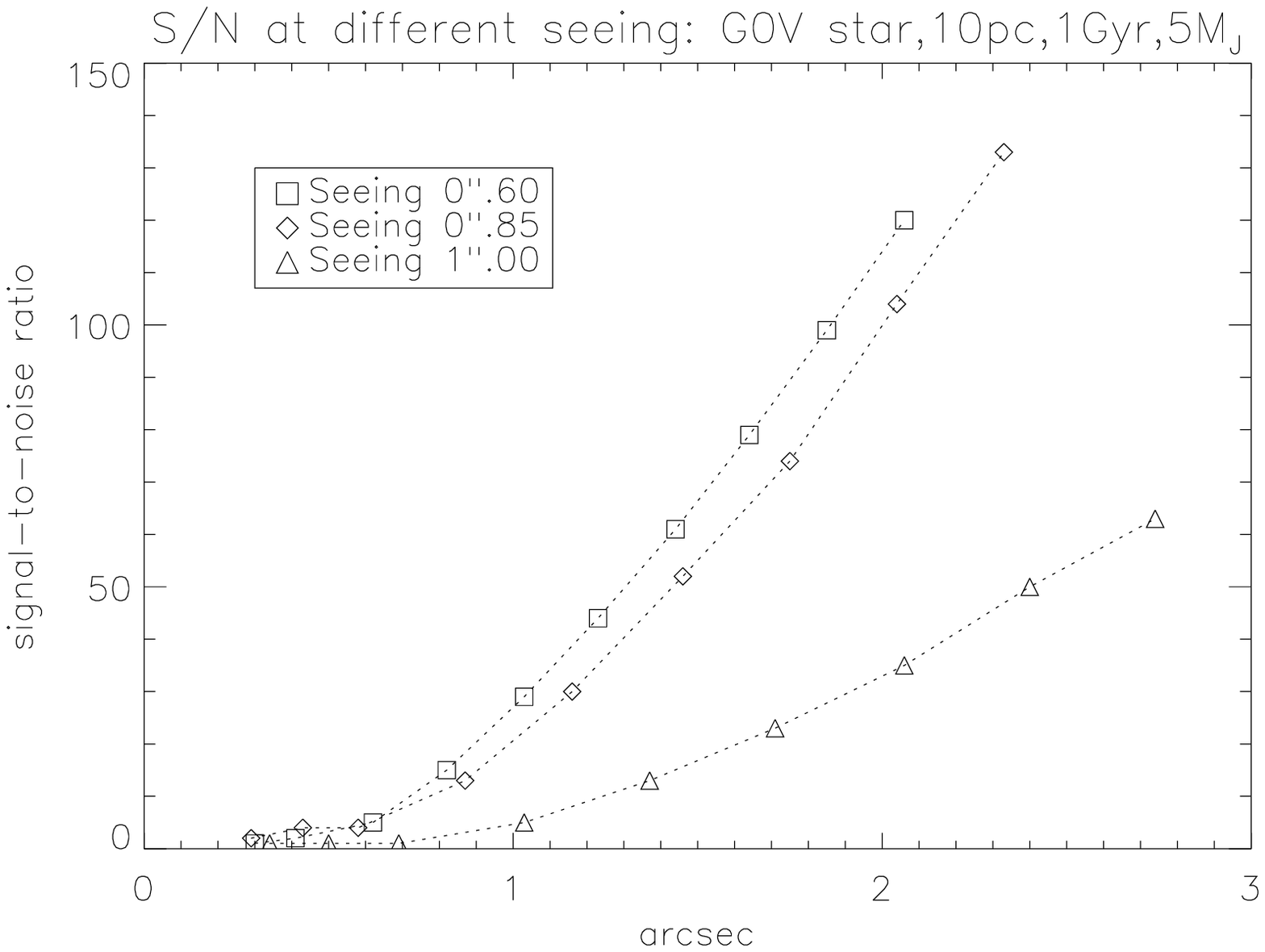}
\plottwo{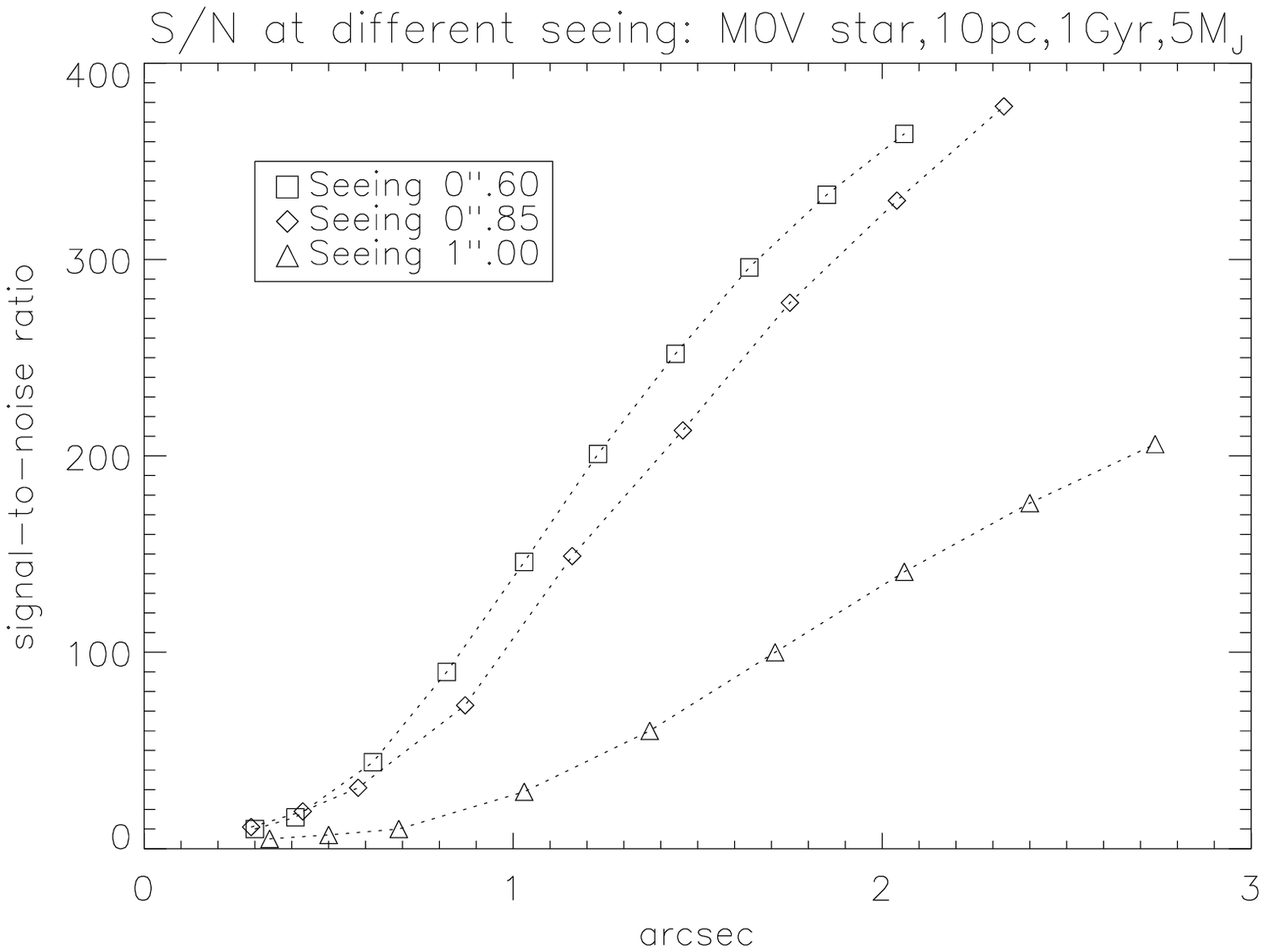}{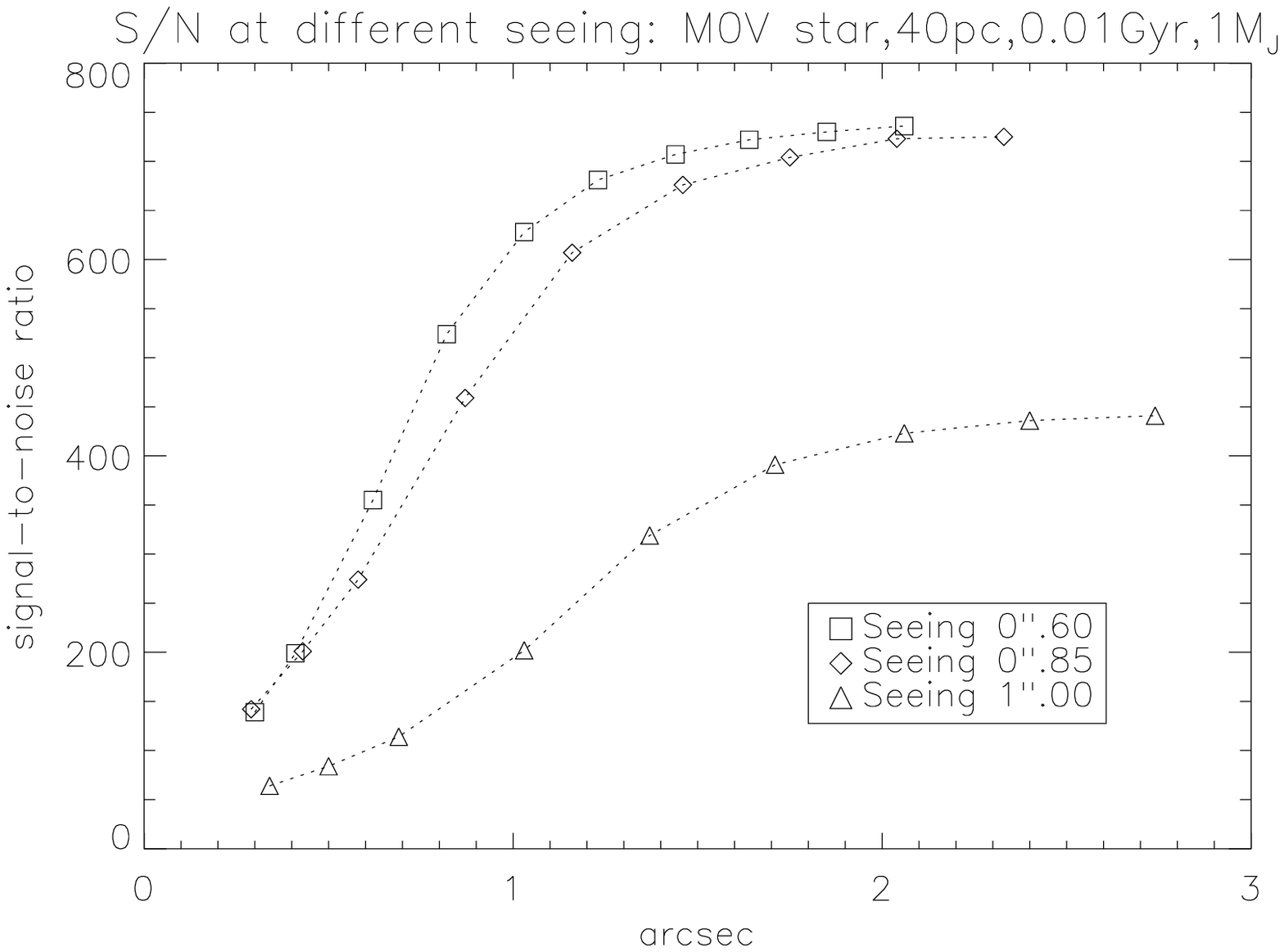}
\caption{Signal-to-noise ratios for detection of four planets at three
  different seeing conditions and at different angular separations. The four
  cases are described in Sec.\ref{sec:seeing}. Every value of seeing have to
  be associated to a Strehl Ratio $S$ as explained in the text: $S$=0.80 for
  the best seeing (0\farcs60), $S$=0.75 for the median (0\farcs80) and
  $S$=0.40 for the worst case (1\farcs00). While the ratios are quite similar
  between $S$=0.80 and $S$=0.75, they decrease dramatically in case of a
  seeing of 1\farcs00, because of the Strehl ratio achievable by the optical
  system, which is only 0.40. \label{fig:seeing}}
\end{figure}

We show here the results for four significant cases:
\begin{itemize}
\item A 5\,M$_J$ planet, 1\,Gyr old, around a G0V star at 3\,pc.
\item A 5\,M$_J$ one, 1\,Gyr old, around a G0V star at 10\,pc.
\item A 5\,M$_J$ one, 1\,Gyr old, around a M0V star at 10\,pc.
\item A 1\,M$_J$ one, 0.01\,Gyr old, around a M0V star at 40\,pc.
\end{itemize}
The resulting signal-to-noise ratios are 
presented graphically in Fig.\ref{fig:seeing}.
The plots show
that the quality of detection decreases as the FWHM increases. This is due to the fact that the Strehl Ratio ($S$) achievable by the AO system is different in the three observing conditions. 
While the achievable signal-to-noise ratio does not change too much when going
from optimal (seeing 0\farcs60, Strehl ratio of 0.80) to good (seeing
0\farcs85, Strehl ratio of 0.75) conditions, planet detection is clearly much
more difficult in the not-so-good condition (seeing 1\farcs00, Strehl ratio of
0.40)\footnote{As explained in Sec.\ref{sec:ao}, our AO simulator is not
  detailed enough to claim a true significance of the dependance of
  signal-to-noise ratio on seeing. Rather, the Strehl numbers themselves
  should be seen as independent parameters.}. 

\begin{figure}
\epsscale{1.0}
\plottwo{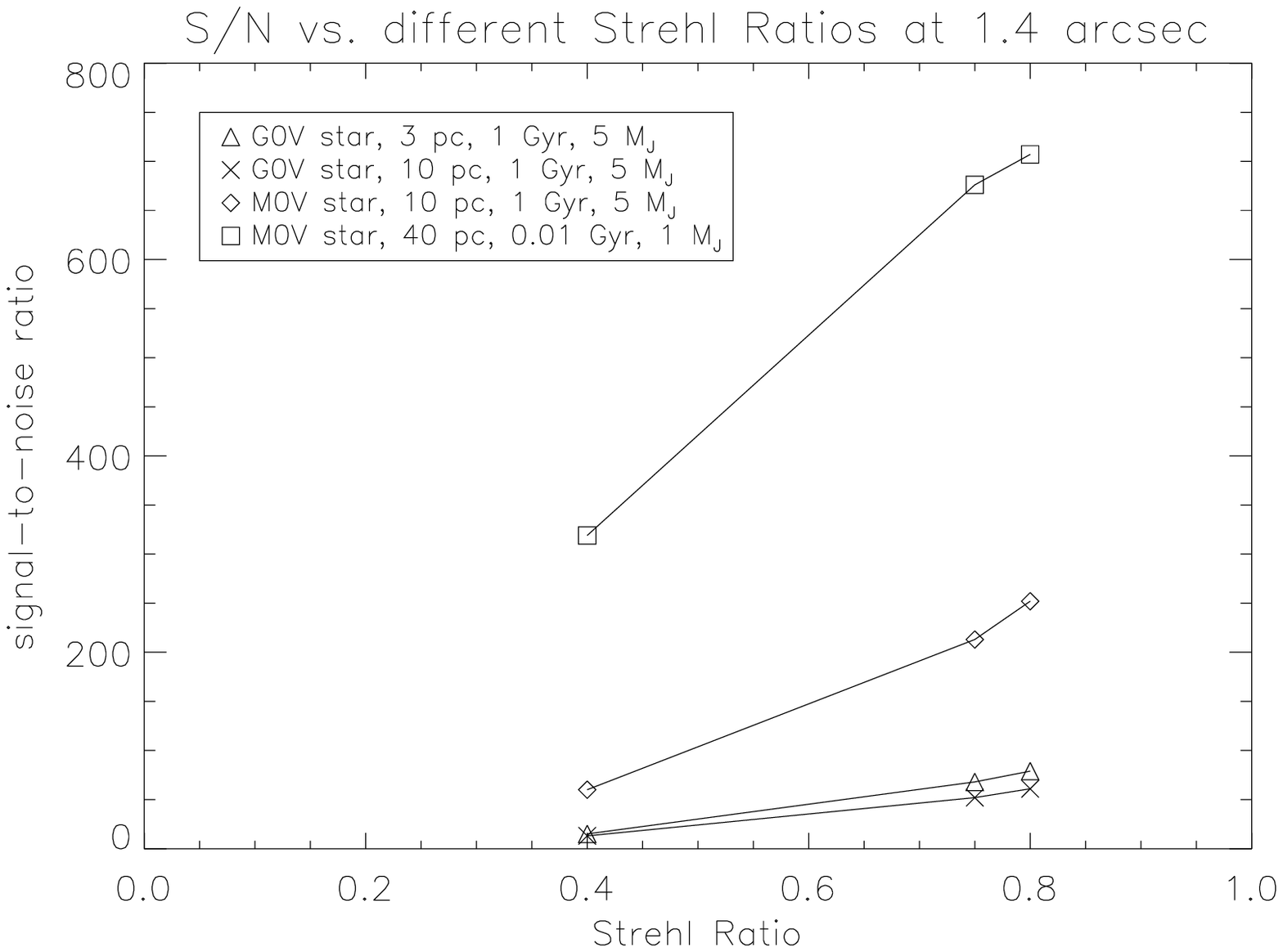}{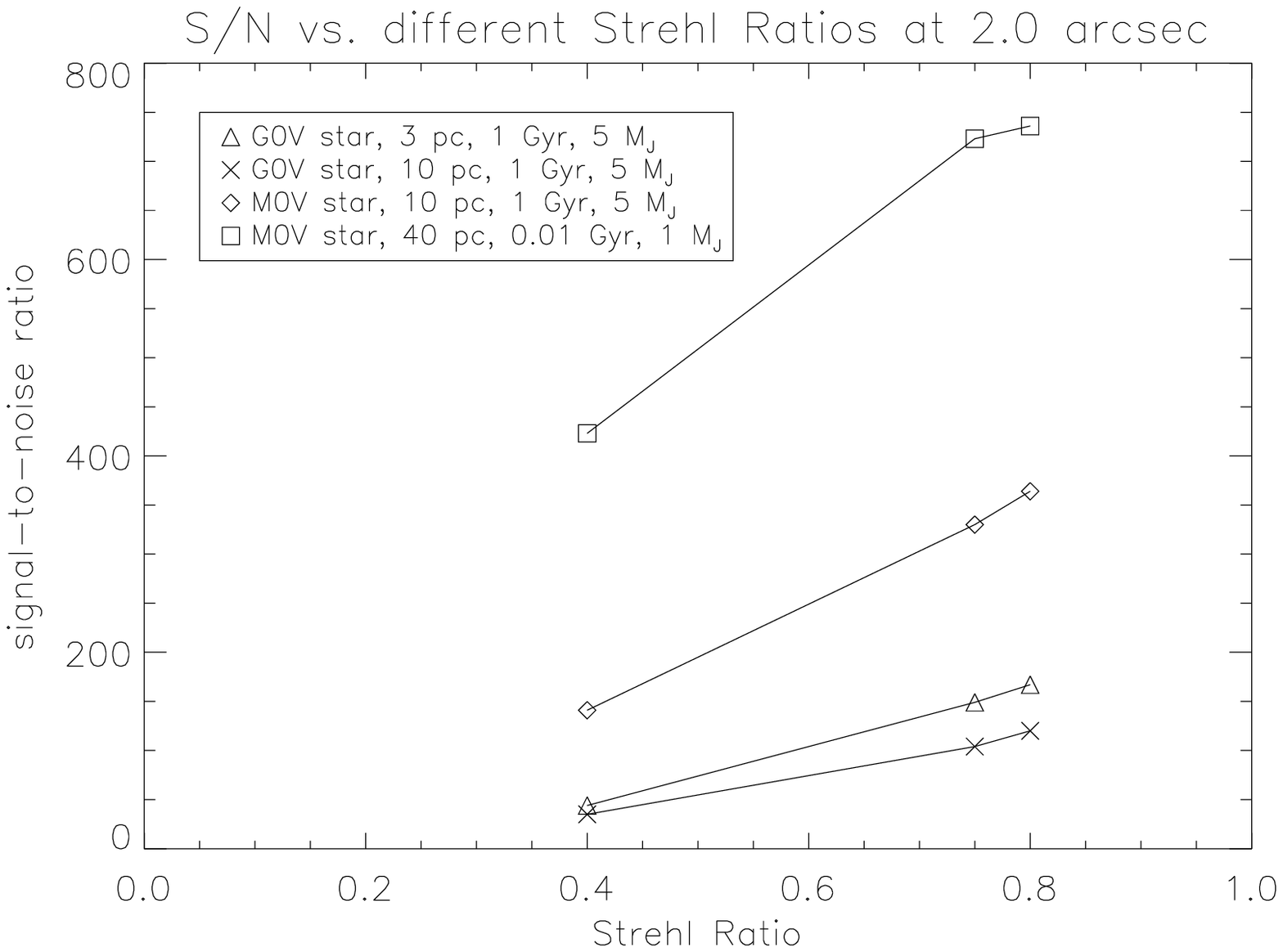}
\caption{Signal-to-noise ratios as a function of the achieved Strehl ratio under the three seeing conditions, measured at 2 angular separations: 1\farcs4 and 2\farcs0. The four planets studied are described in Sec.\ref{sec:seeing}. The Strehl Ratios are stronlgy dependent on the seeing and on the capabilities of the AO system. \label{fig:sr}}
\end{figure}

In Fig.\ref{fig:sr} the signal-to-noise ratio is presented as a function of Strehl ratio for two angular separations. The signal to noise ratios, going from $S$=0.80 to $S$=0.75 and $S$=0.40, are reduced by a factor of $\sim$2. Since the value of $S$ depends not only
on the atmospheric turbulence and seeing but also on the performances of the AO systems,
these plots show in a more general way how important is to work both with a good seeing and with a very efficient AO to detect extrasolar planets.

\section{Discussion}
\label{sec:discussion}

\begin{figure}
\epsscale{.50}
\plotone{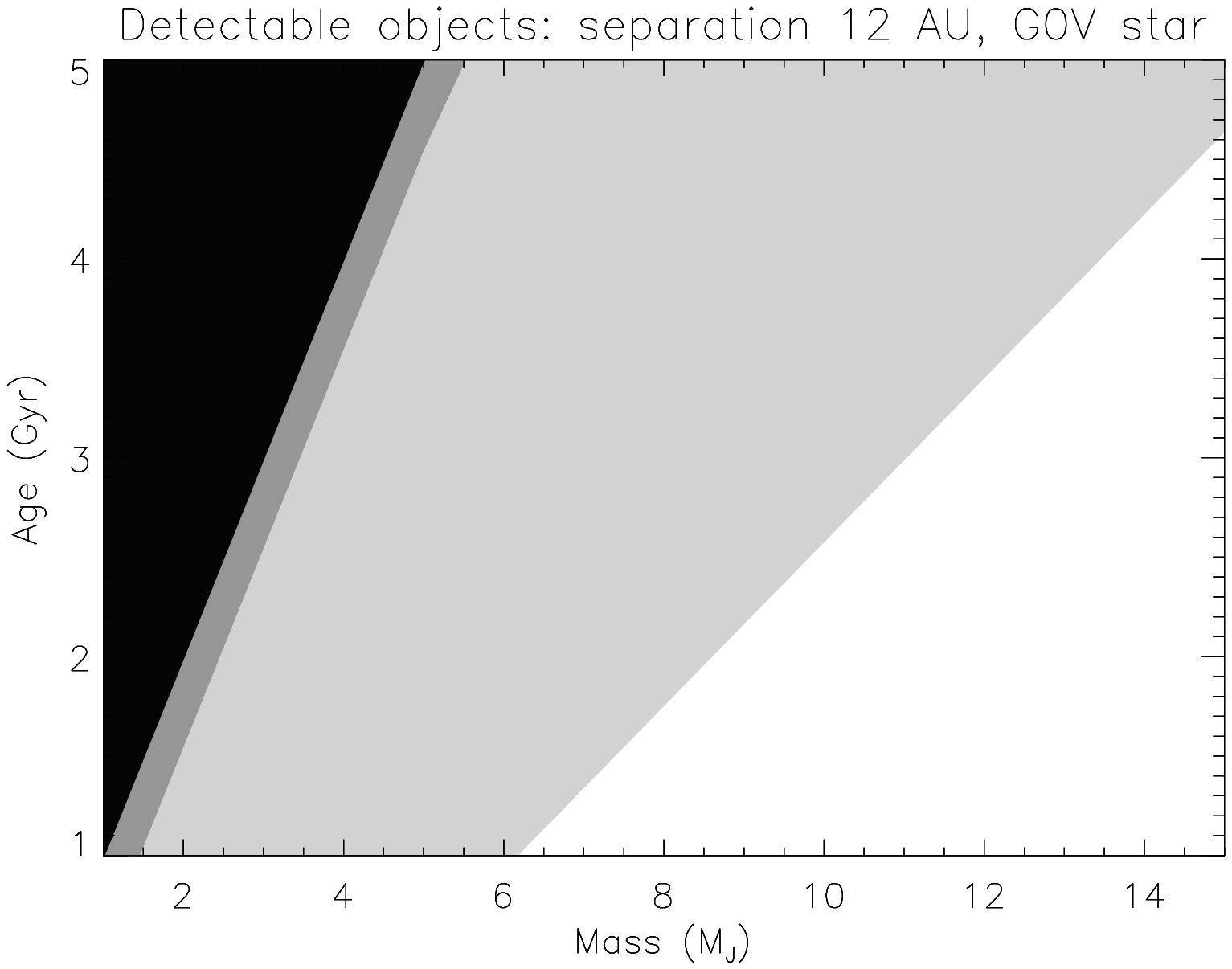}
\plotone{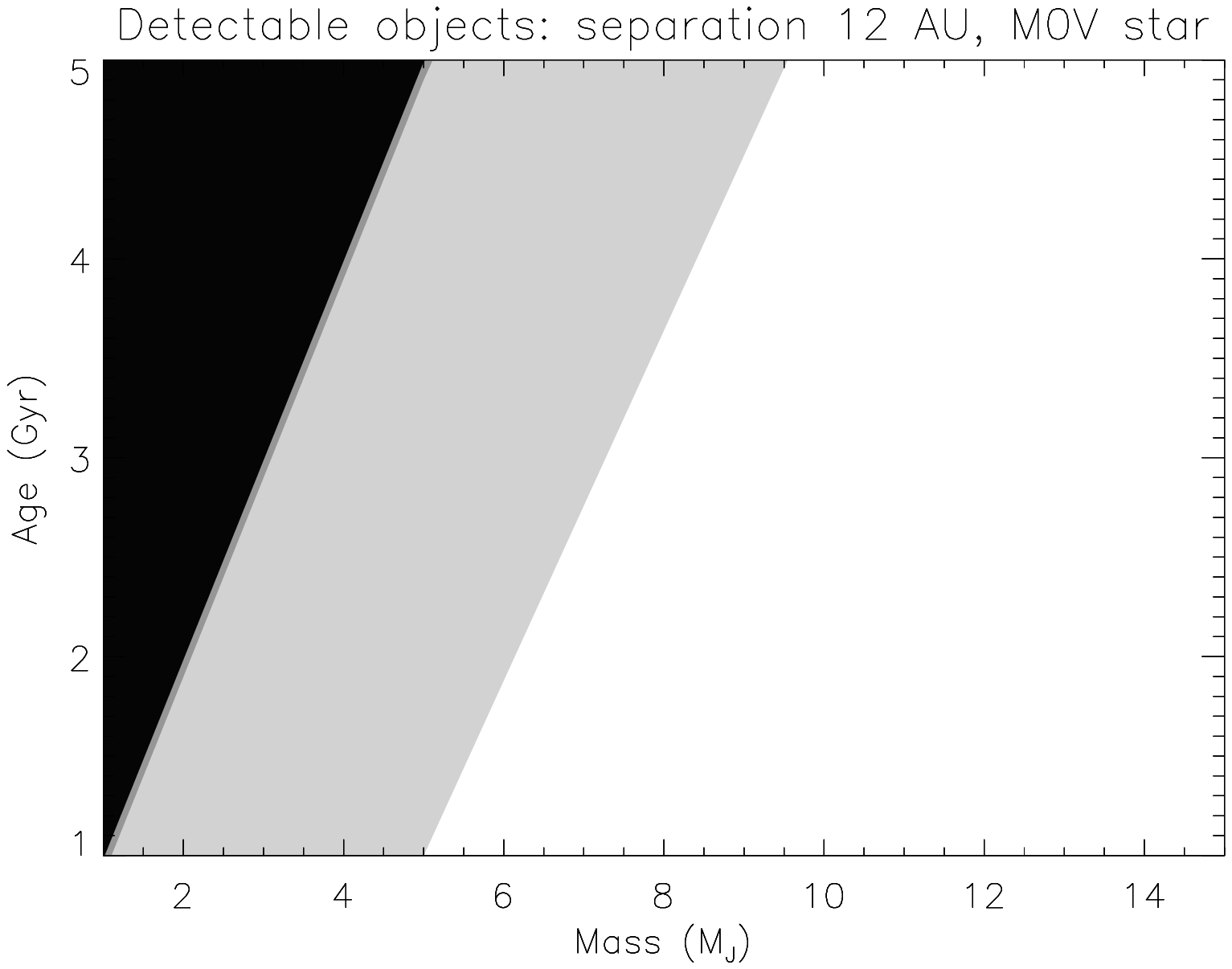}
\caption{Top: planets detectable with a signal-to-noise ratio higher than 5, as function of mass and age, in the case of a G0V host star. The most massive objects (30\,M$_J$) and the youngest planets (0.01\,Gyr) are not included in the plot because they can always be detected. The white region represents the planets always detectable, for all the distances considered (3, 10 and 40\,pc). In the light-gray region there are the planets not detectable if they are at 40\,pc from the Sun, but detectable at 10 and 3\,pc. In the dark-grey region the planet are not detectable if they are located at 10\,pc or further away, but they are visible at 3\,pc. In the black region planets are detectable at none of the considered distances. The physical separation between the star and the planet has been kept at 12 AU, but this is changing the angular separations, which are respectively 4\farcs0 at 3\,pc, 1\farcs2 at 10\,pc and 0\farcs3 at 40\,pc. Bottom: the same plot as before calculated for an M0V star. Both these plots have been obtained using data from Tab.\ref{tab:plotg}. \label{fig:prova}}
\end{figure}

From the data shown in Tab.\ref{tab:plotg} we can see that most of the targets we
simulated to observe are actually detectable. In particular, it is clear that
very massive objects (30\,M$_J$) and very young ones (0.01\,Gyr) are detected at every considered distance with very high signal-to-noise ratios
(usually higher than 1\ 000). In the other cases, the signal-to-noise ratios are strongly dependent on the age and on the mass of the planet, as shown in
Fig.\ref{fig:prova}. From these plots we can conclude that the
detection is easier in the case of larger masses, or, if the mass is fixed, for
young objects.

The less massive objects (from 1 to 5\,M$_J$) are more interesting targets, but they are also much more difficult to detect. Their potential detection is more sensitive to the
distance of the planetary system (they are almost unobservable at 40\,pc)
because of their faintness in the near-infrared region. In some cases they are shining by reflected light, namely the 1\,M$_J$ planets older
than 1\,Gyr, but also the 5\,M$_J$ 5\,Gyr old. Systems with planets shining by reflected light are the most interesting cases,
because of the similarities with the Solar System. Detection of 
a planet like Jupiter (1\,M$_J$, 5\,Gyr old) around a star almost like the Sun, 
at a separation around 5 AU, means to observe something so similar to
our system to allow us to suppose the presence of terrestrial planets in
inner orbits, perhaps even in the habitable zone. All reflected light cases
simulated are not detectable with 4 hours of exposure time, even if they
were only at a distance of 3\,pc from the Sun. This kind of object is still a challenging goal
for such an instrument, at least if it is mounted on an 8m-class telescope. They can be better detected using different strategies dedicated to reflected light, for example the
CHEOPS project with the polarization analyzer ZIMPOL
.

From the plots in Fig.\ref{fig:prova} we can also conclude that the SDI-IFS technique allows detection of planets of a few
Jupiter masses (around 5), with ages of $\sim$1\,Gyr, within a distance of 10\,pc. Passing from 10 to 3\,pc we are not enlarging dramatically the range of planets detectable, as shown by the very thin dark-gray region in both plots. Planets of a few Jupiter masses, up to now, have been directly detected only if they are very young, while
we can consider a 1\,Gyr planet to be reasonably old. The direct detection of
such an object would be a very important result in the present contest of
searching for extrasolar planets.

Finally, comparing the two plots, we note that around an M star it is
possible to detect more objects at a distance larger than 10\,pc. At 40\,pc in
this case we can detect some interesting old planets down to $\sim$10\,M$_J$.
The reason of these better detections for an M star are simply related to the better contrast due to the lower brightness of the M star with respect to a G
star at the same distance. The spectral differences between the two stars at
this resolution are less important than the differences in flux. Less flux
means less photon noise and less speckle noise.
In the case of less flux, beside an improvement of the signal-to-noise ratio, we have also an increment in importance of the photon noise of the planet, that becomes
predominant in case of very bright planets and at large angular separation of the star: this can be seen, for example, in Fig.\ref{fig:seeing}, bottom right,
when the signal-to-noise ratio is stabilizing at a given level beyond $\sim$2
arcseconds.


Although these results showed that many detections are possible,
if we calculate the real probability to find planets in the solar
neighborhood we have to
notice that the list of observable targets is not very long, because in  
the considered ranges of distances (3, 10 and 40\,pc) there
are only few stars that can harbour planets which are at the
same time detectable with our IFS-SDI.
With data from the Hipparcos catalog, \citet{des}
considered a sample of more than 650 stars within 20\,pc,
evaluating spectral types, ages, metallicities and binarity; 
for realistic but rather conservative assumptions on the period distribution of giant planets extrapolated from the results of RV surveys, it was concluded that the detection of at least 10 giant planets is actually expected.
Half of the chance of finding planets is confined to the best $\sim$80
targets, while with more than 200 we have almost no chances of success. Shrinking
the distance
to $\sim$3\,pc, for which we obtained from simulations the better
detections,
we reduce the probability to negligible values.

It should not be forgotten that all these results and considerations are related to the case of an 8m-class telescope. It is natural to expect very 
different results for different instrumental parameters. We suppose that the
detection of faint planets, specially those shining by reflecting light, would be significantly easier with a larger telescope, in particular one of
the new Extremely Large Telescopes (ELTs) on project, which cover a
range of diameters between 20 and 100 m. The structural and technical 
differences between an ELT and a 8m-class telescope are sometimes not easy
and immediate to consider (e.g. segmentations, spiders, AO systems, etc.), so
we need new sets of specific simulations to quantify the improvement in
signal-to-noise ratios passing from an 8 meter telescope to an ELT. These
simulations will be presented in a forthcoming paper.


\subsection{Advantages in using IFS-SDI}
\label{sec:why}

The reasons why an SDI system based on an IFS is preferable to a set of filters can be
summarized in three crucial points:
\begin{itemize}
\item First of all, with a whole spectrum available we are not forced to study
only one spectral feature of the planet. If we have two filters, or three,
chosen to observe a specific absorption band, we lose all information coming
from the other features. With a spectrum we can actually combine the data
on and off different absorption bands to increase the signal-to-noise ratio of
our detection (if there are other good features).
\item Second, the availability of data at many wavelengths can allow the
development of various algorithms and specific methods for data reduction 
to improve the reduction in speckle noise. 
\item Finally, the most important reason for using IFS is the possibility of
not only detecting extrasolar planets, but also characterizing these objects.
In fact, the data available after the subtraction of stellar light allow the
study of the spectrum of the planet and its main features, which can be very
different according to mass, age, temperature etc. This argument is 
fundamental to understand that our method can be very useful also in 
observing planets already discovered in other ways. 
\end{itemize}

Another potential advantage of an IFS, with respect to a conventional dual
band imager is that one can actually make an all-reflective design up to the
point where the image is formed, minimizing chromatic errors (as already said
in Sec.\ref{sec:code}).

We present in this section a test made on the simulated images in order to
verify the first two points expressed above, that means the actual gain in
signal-to-noise ratio and noise reduction due to analysis of a number of
wavelengths larger than 2 or 3. To do this, we applied four different
operations to the monochromatic images obtained from the spectra, accurately
rescaled to a given wavelength. The four operations combine
2, 3, 6 and 9 different resolution elements. The images are indicated by $I_n$.

\begin{enumerate}
\item Single difference, with 2 elements: $I_2 - I_1$.
\item Double difference, with 3 elements, which is Eq.\ref{eq:dd}.
\item 3$^{rd}$ difference, with 6 elements, is given by
\begin{eqnarray}
\lefteqn{ DDD = \left[ \left( I_1 - I_2 \right) - k_1 \left(I_1 - I_3 \right) \right] +} \nonumber \\
& & -K \left[ \left( I_4 - I_5 \right) - k_2 \left(I_4 - I_6 \right) \right].
\end{eqnarray}
\item The 4$^{th}$ difference is a combination of two 3$^{rd}$ differences:
{\setlength\arraycolsep{2pt}
\begin{eqnarray}
DDDD & = & DDD_1 - \kappa DDD_2 ={}  \nonumber \\
& & {}\left[ \left( I_1 - I_2 \right) - k_1 \left( I_1 - I_3 \right) \right] + {}\nonumber \\
& & {}-K_1 \left[ \left( I_4 - I_5 \right) - k_2 \left(I_4 - I_6 \right) \right] + {} \nonumber \\
& & {}- \kappa \left[ \left( I_4 - I_5 \right) - k_3 \left(I_4 - I_6 \right) \right] + {}\nonumber \\
& & {}+ \kappa K_2 \left[ \left( I_7 - I_8 \right) - k_4 \left(I_7 - I_9 \right) \right] {}.
\end{eqnarray}}
\end{enumerate}

\noindent The values $k_n$, $K_n$ and $\kappa$ are constants that have to be 
defined case by case in order to subtract the speckles and their residuals as best as possible. Every subtraction,
other than reducing speckle noise, introduces a slight increment of the other
sources of noise (photons, sky background, detector noises etc.) that has to be considered carefully. 

\begin{figure}
\epsscale{.70}
\plotone{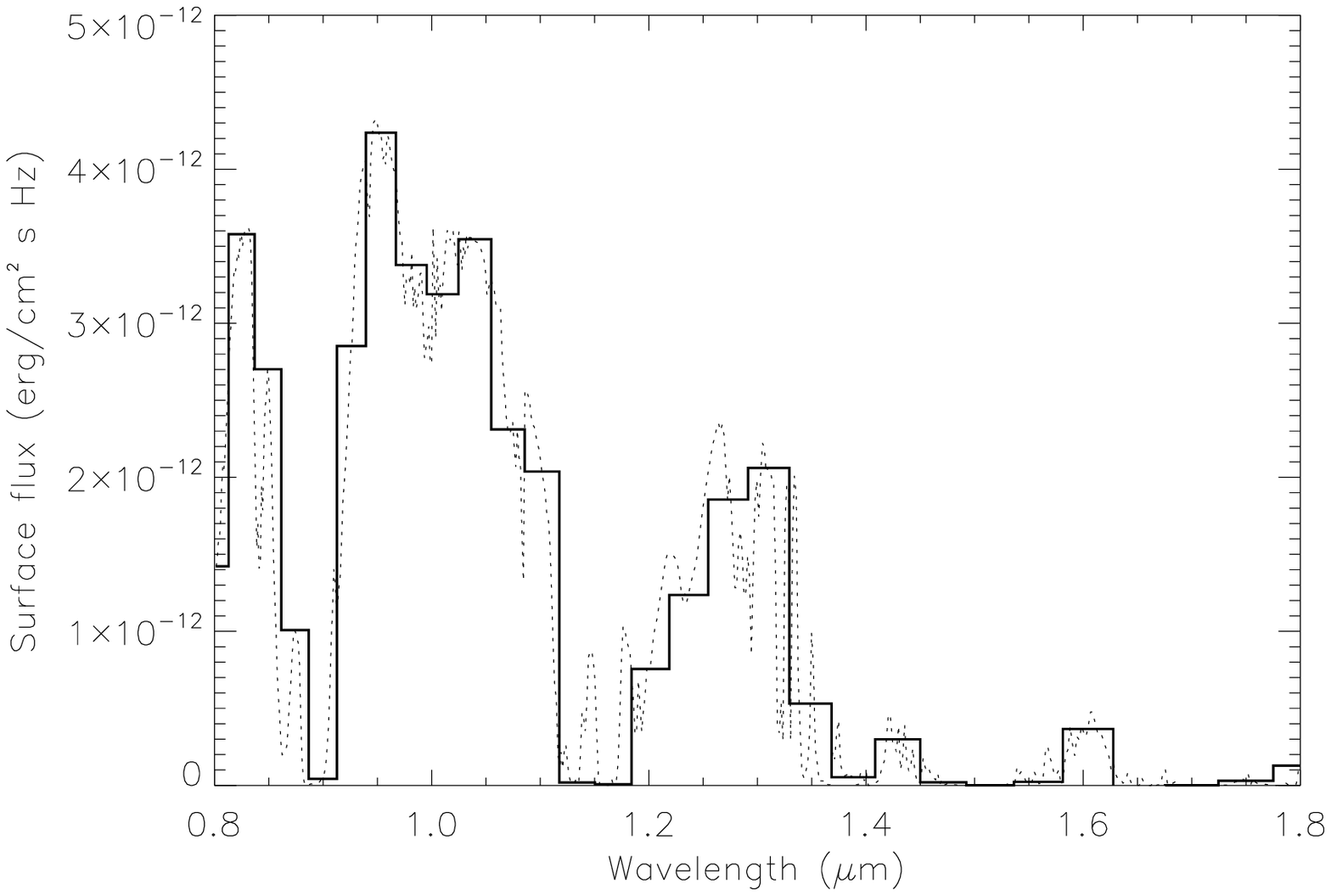}
\plotone{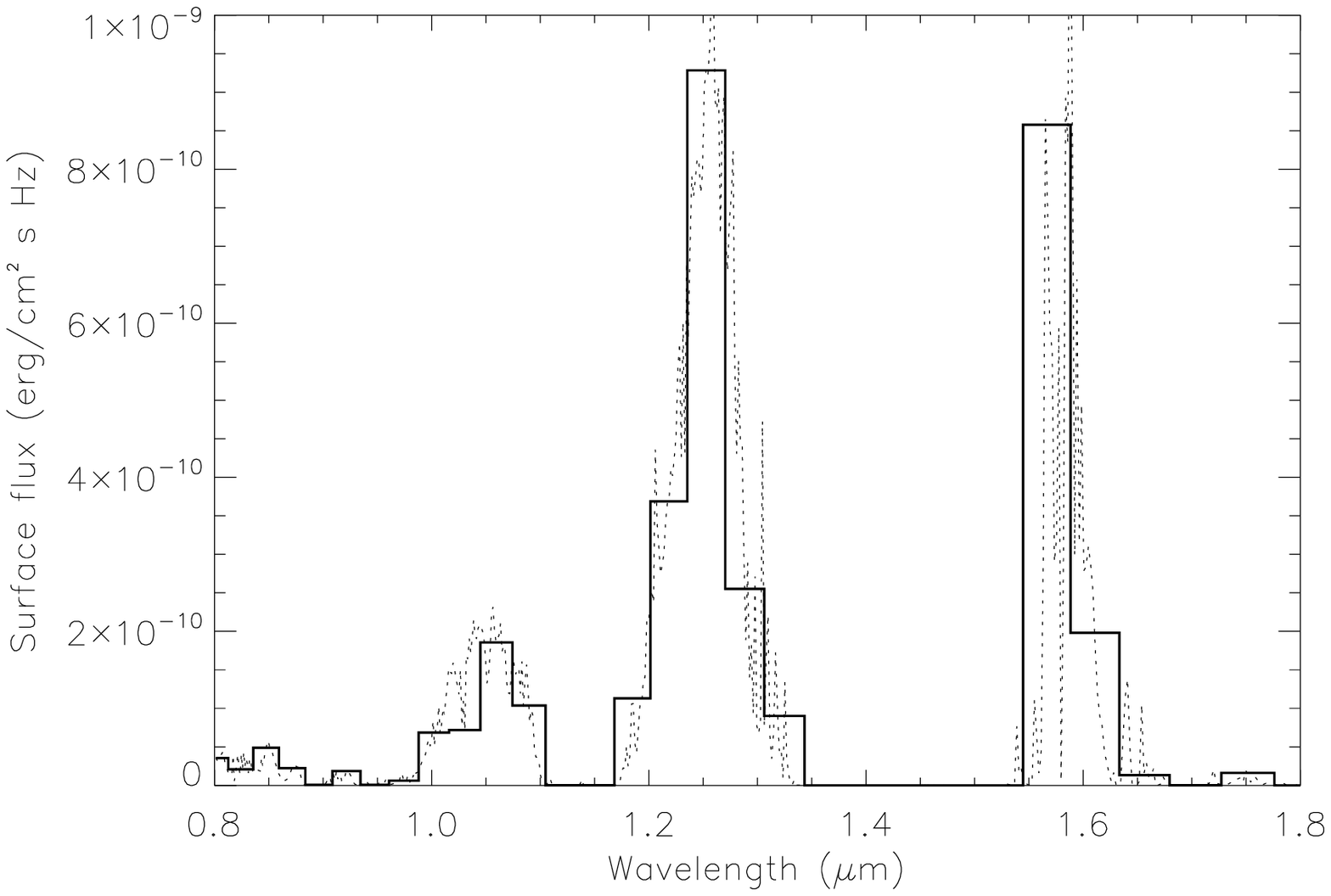}
\caption{Top: Near-infrared spectrum (between 0.8 and 1.8$\mu$m) of a 1\,M$_J$
  planet orbiting a solar-type star (G2V) 1\,Gyr old, with a separation of 10 AU. Bottom: The spectrum of the same planet but at a separation of 1AU. For both these planets the spectral emission is mainly intrinsic, with a negligible component of reflected light. The solid spectrum has a spectral resolution of 30, the dotted line of 750. Data taken from \citet{bur}. \label{fig:plan}}
\end{figure}

At first, we measured the total level of noise in a specific region of the
stellar PSF normally dominated by speckles (in our case, for an 8m-class telescope,
we worked between 0.4 and 0.6 arcseconds, avoiding spiders and artifacts). We used different simulations of the
same object, and we obtained averaged values of residual noise in this region
using all the four different algorithms. All these calculations of noise have
been done without planets.
We calculated then the signal of a planet, chosen bright enough to be 
detectable clearly (signal-to-noise ratio $\sim$5) with the single difference 
method, and with all but one of the resolution elements considered in absorption
bands: practically, the signal of the planet is significant only in one of
the nine cases. In this way, we simulated a detection
using nine different wavelengths but only one of the three ``peaks'' of the
planetary spectrum shown in Fig.\ref{fig:plan}. We calculated the 
signal-to-noise ratios of the detection, as a function of the operation used
for subtracting speckle noise.

\begin{figure}
\epsscale{.70}
\plotone{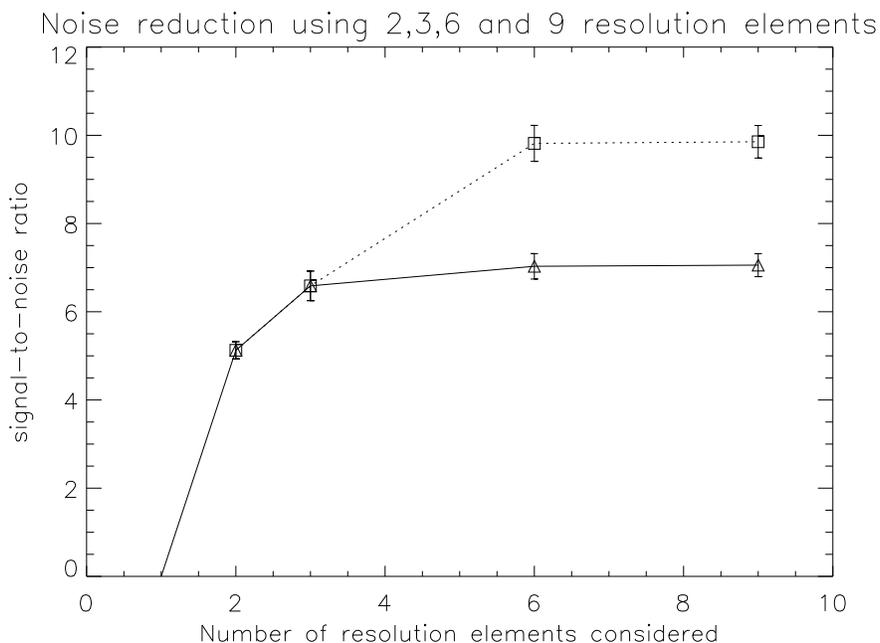}
\caption{Improvement of signal-to-noise ratio of a detection using combinations of different numbers of resolution elements. The triangles connected by the solid line are the results obtained considering a single planetary spectral features; the squares, connected by a dotted line, are obtained considering simultaneously two similar features, which is possible only with IFS-SDI and only considering more than 3 resolution elements. More details are explained in the text. The noise level for this plot has been calculated in various regions of the images and using various simulations of the same object, so the result shown are averaged signal-to-noise ratios with error bars given by the standard deviations. Although the improvement in detection is not very dramatic using only one spectral feature and a number of resolution elements larger than 3, the possibility of using two spectral features instead of one can be really significant, as shown by the dotted line in the plot. \label{fig:f}}
\end{figure}

In a second step, we tried to elaborate the results a little more in order to
also use the other ``peaks'' of the spectrum of the planet. Practically, 
considering that the signal of the planet in SDI is given by the contrast in
flux of the planet on and off an absorption band, we combined the contrasts in
two of the three peaks to increase the signal-to-noise ratio. We used the spectrum of a 1\,M$_J$ planet at 1AU from a solar-type 
star (Fig.\ref{fig:plan} at bottom), where we have two peaks of very similar maximum intensity. The results are shown in Fig.\ref{fig:f}.


It appears obvious from this plot that a number of resolution elements larger than
3 can actually improve the signal-to-noise ratio of the detection. In fact
the improvement is significant, but not dramatic, because just after the double
difference the total noise is less dominated by speckles and more by photons.
The photon noise represents a limit that cannot be over passed just combining
more elements. Moreover, with 9 resolution
elements the signal-to-noise ratio doesn't increase.
To conclude, a combination of six elements seems to be the best solution for
reducing speckle noise, keeping all the other sources of noise under control.

On the other hand, the simultaneous use of other planetary features, 
represented by the three ``peaks'' of Fig.\ref{fig:plan} at bottom, provides
an interesting improvement of the signal of the planet, and consequently of the
signal-to-noise ratio. Actually, we used only the two highest peaks at 
$\sim$1.25 and $\sim$1.60$\mu$m, because the contribution of the third peak (at
$\sim$1.05$\mu$m) is negligible if compared with the increment of noise 
combining it with the other data. As shown in Fig.\ref{fig:f}, we actually
improved the signal-to-noise ratio by a factor of $\sim$$\sqrt{2}$, because the
contrast for both peaks is very similar. 
Since a dual band imager can enhance some specific spectral features depending
on the used filters, it is not easily achievable to observe at the same time
many of these different features, while to have all the peaks in a single
simultaneous exposure is possible with IFS.


\section{Conclusion}

With numerical simulations we evaluated the capabilities of an IFS
instrument to reduce speckle noise. One can achieve a reduction factor of 
10$^{-3}$, which can be considered a good
result for the SDI technique. It allows the detection of a vaste range of
faint companions around stars in the solar neighbourhood. For an 8m-class
telescope, our simulations have shown that objects of a few Jupiter masses can
be clearly detected within 10\,pc, with very high signal-to-noise for the most massive and youngest ones. Much more difficult will be the detection of old and
cold Jupiter-like planets shining by reflected light. From these first simulations it seems that such an instrument mounted on an 8 meter telescope cannot
detect planets like Jupiter in a system like our Solar System, at least with exposures of 4 hours or shorter.
Despite this the IFS method has some very important advantages that should not be neglected, in particular the possibility to characterize
the planet by studying its spectrum. This IFS-SDI is not only useful for planet detection, but in cases of relatively high
signal-to-noise ratio we can extract its whole near-infrared low resolution spectrum. This 
characterization is not completely possible in narrow band SDI systems working with only two or three defined filters, while it becomes possible for SDI-IFS
because of the large spectral information available. This
technique appears to be extremely useful, if not for detecting very cold and
small objects, for a better detection and characterization of planets younger
than 1\,Gyr or more massive than a few Jupiter masses.
The structure of the simulation code, which is versatile and adaptable
to different kinds of instruments, will also allow simulation of
telescopes larger than 8 meters (ELTs). We expected to have a significant
improvement of the quality of detection in case of diameters of 20, 50 or
100m; and we expect, in that case, that also young and cold planets can be
detectable with short exposure times: this will be the topic of a forthcoming paper.

\acknowledgements
This work started during the phase A of the CHEOPS project for a Planet Finder
instrument at VLT on Cerro Paranal (Chile). It was supported by the European
Southern Observatory (ESO). The authors wish to thank all the people involved
in the CHEOPS project for the useful collaboration during this phase
A. A. Berton wants to thank Kjetil Dohlen from the Laboratoire d'Astrophysique de Marseille for his important suggestions and comments, and Catherine Heymans from the University of British Columbia, Vancouver, for her kind help.

\appendix

\section{Flux contrasts for the considered cases}

\begin{figure}
\epsscale{1.1}
\plottwo{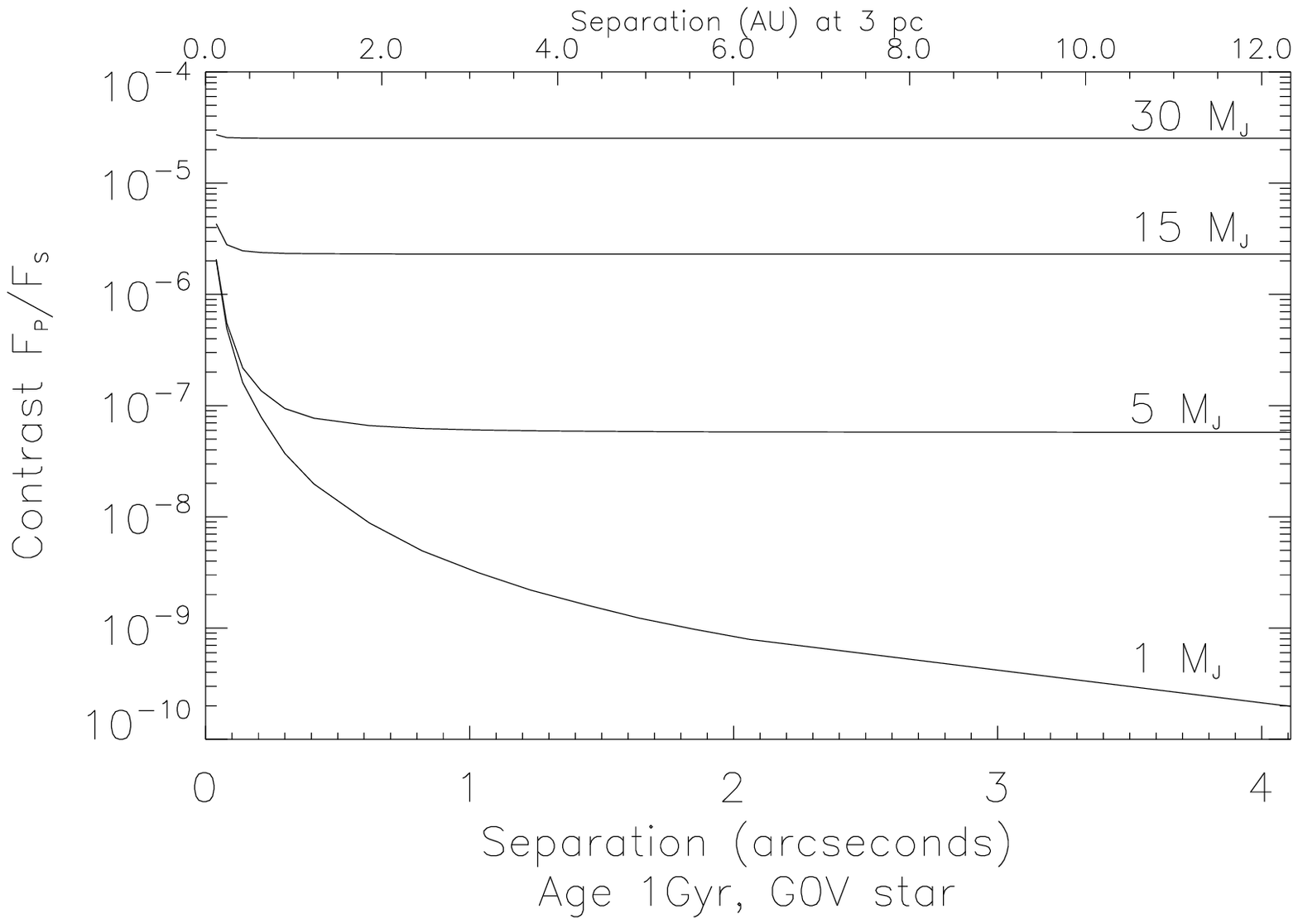}{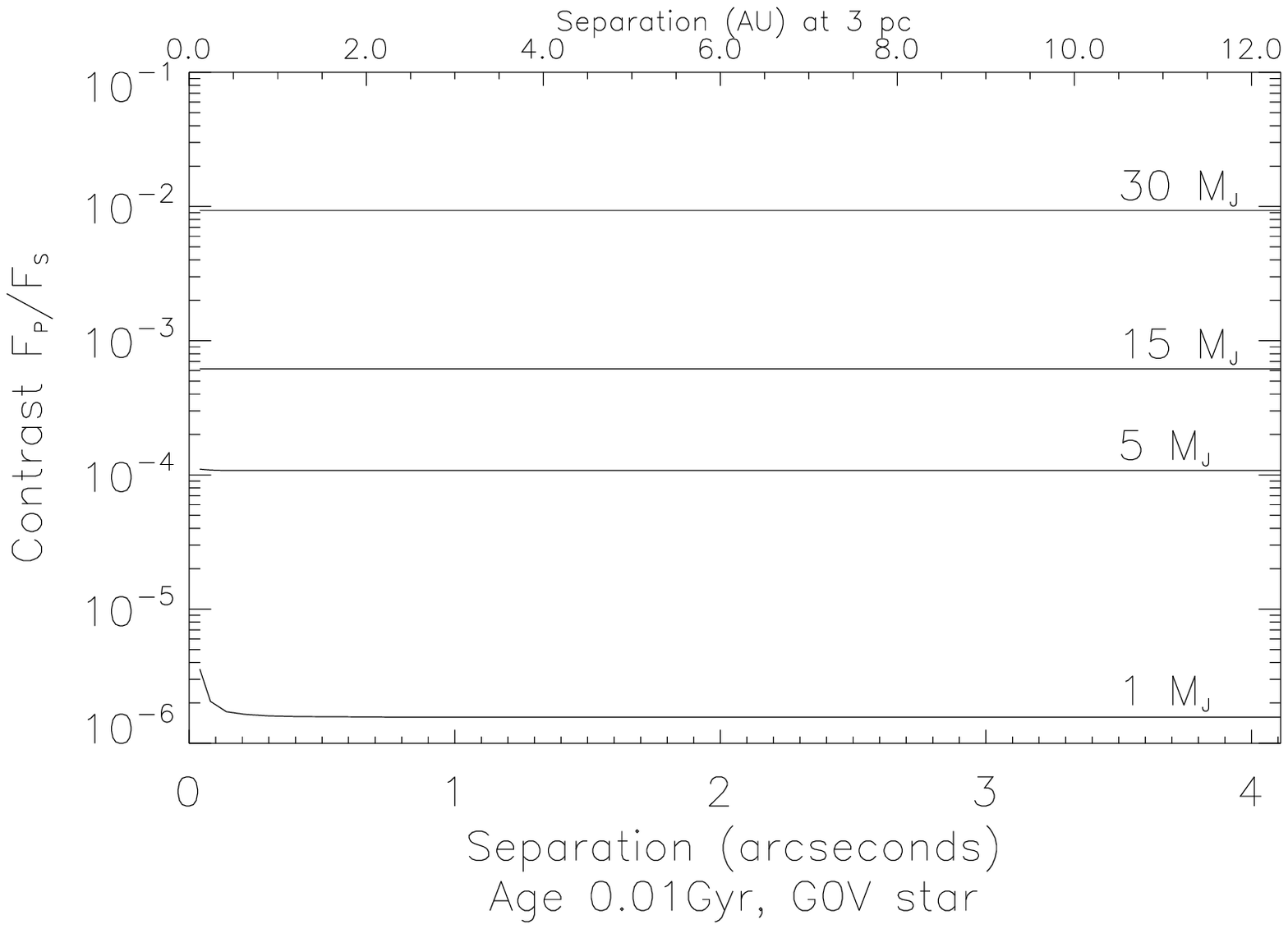}
\plottwo{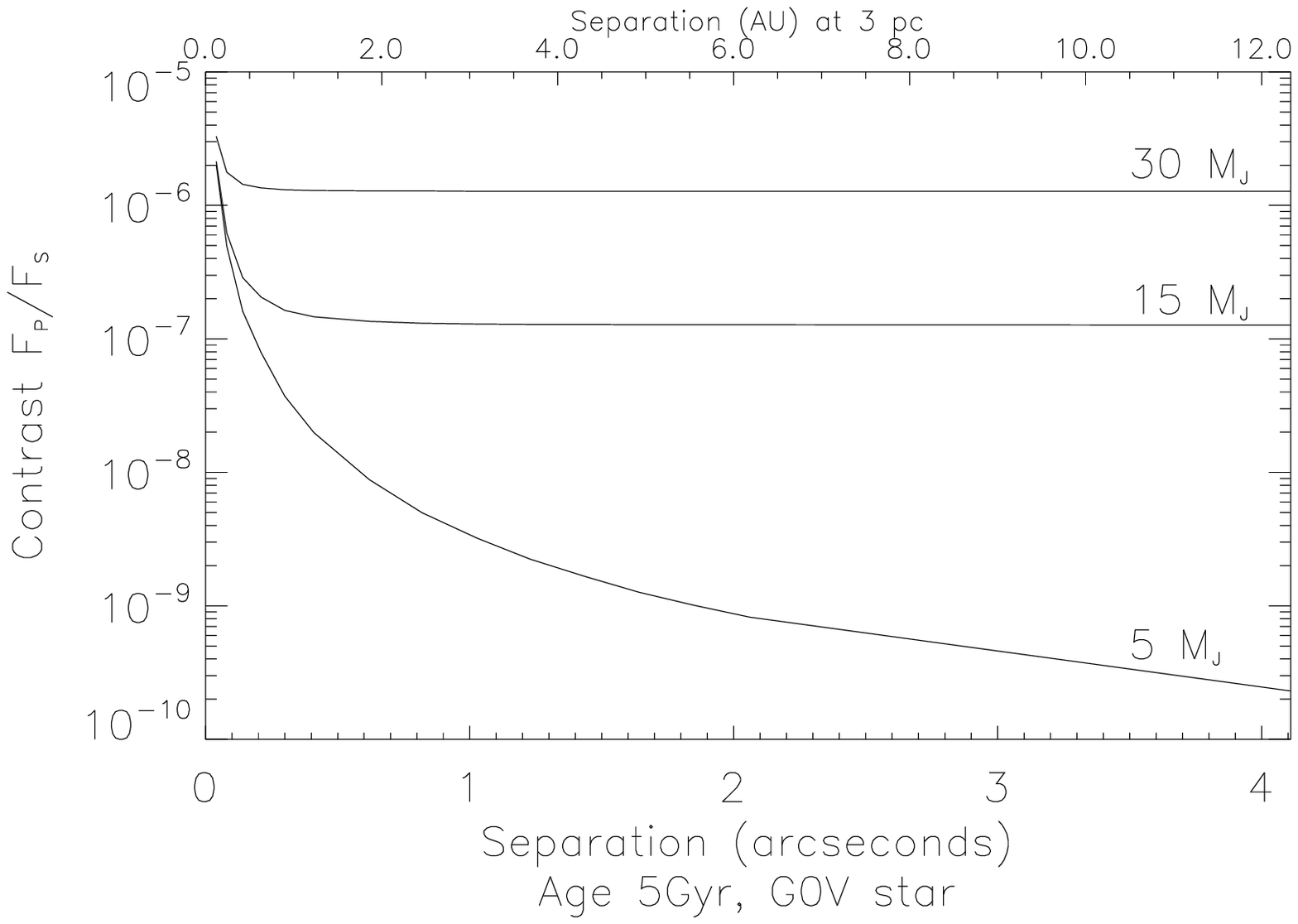}{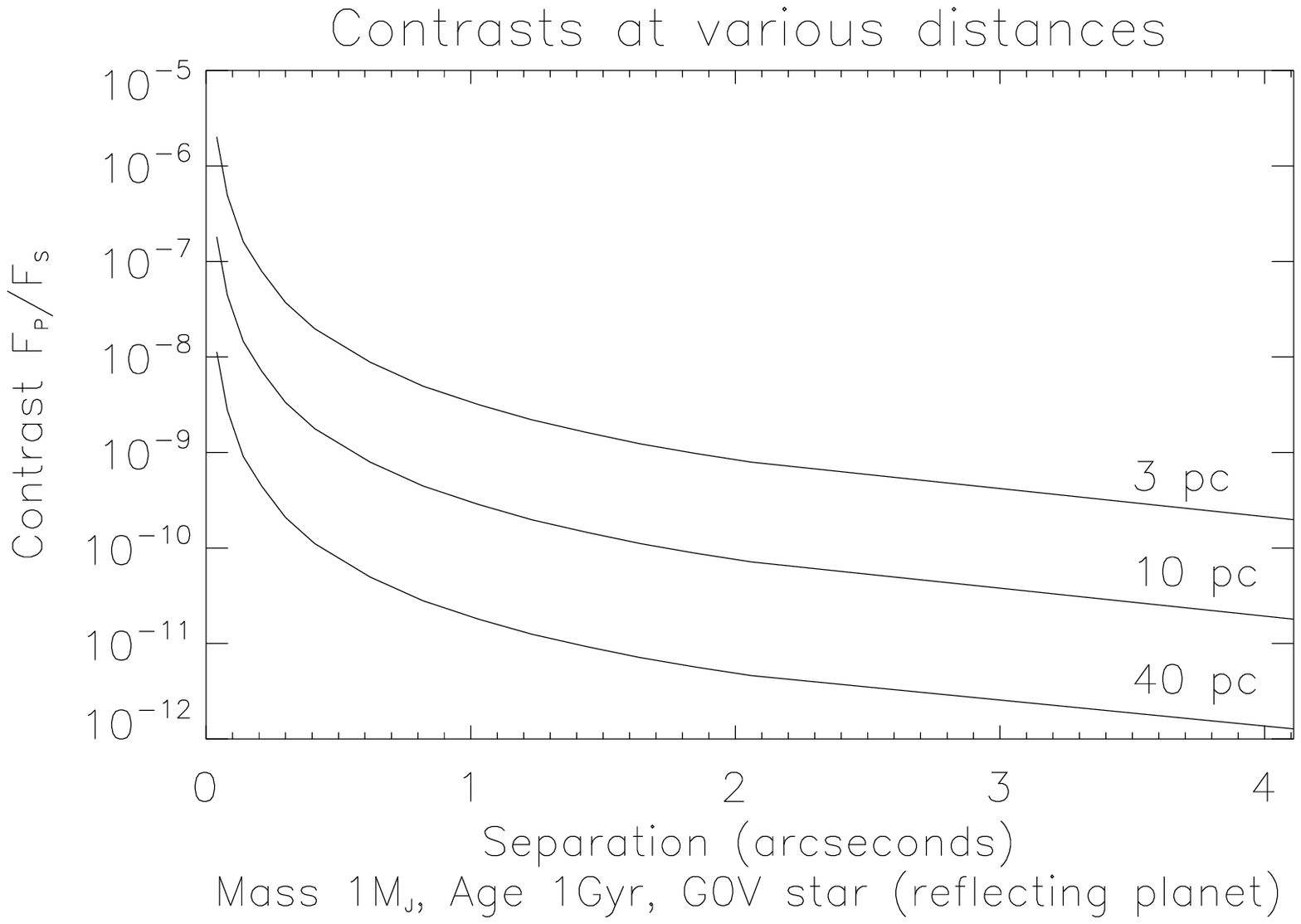}
\caption{Contrasts between the flux of the central peak of a planet ($F_P$) and the
  flux of the central peak of the
  host star ($F_S$) as function of the angular separation, calculated in $J$ band for the planets simulated in
  Sec.\ref{sec:results}. The star is always a G0V star and the distance 3 pc, the ages of the system
  are 1 Gyr (top-left), 10 Myr (top-right) and 5 Gyr (bottom-left). In each
  plot the results are computed for four different masses (1, 5, 15, 30
  M$_J$). The physical separation in AU is also reported. At bottom-right: the
  contrasts as function of the angular separations for a 1 M$_J$, 1 Gyr old
  planet in a system at 3, 10 and 40 pc from the Sun. Data taken from
  \cite{bar}. \label{fig:contrasts}}
\end{figure}

\begin{figure}
\epsscale{1.1}
\plottwo{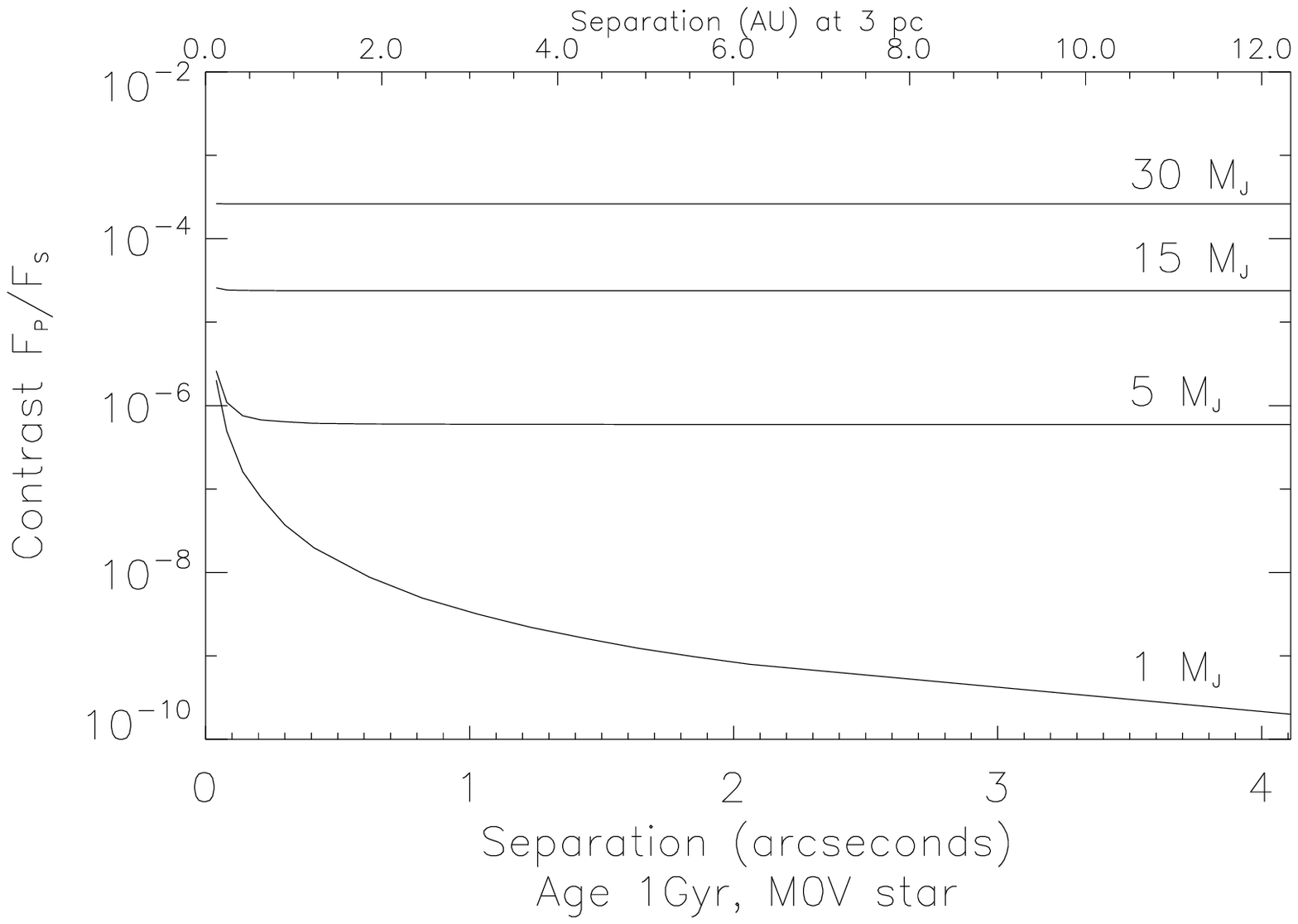}{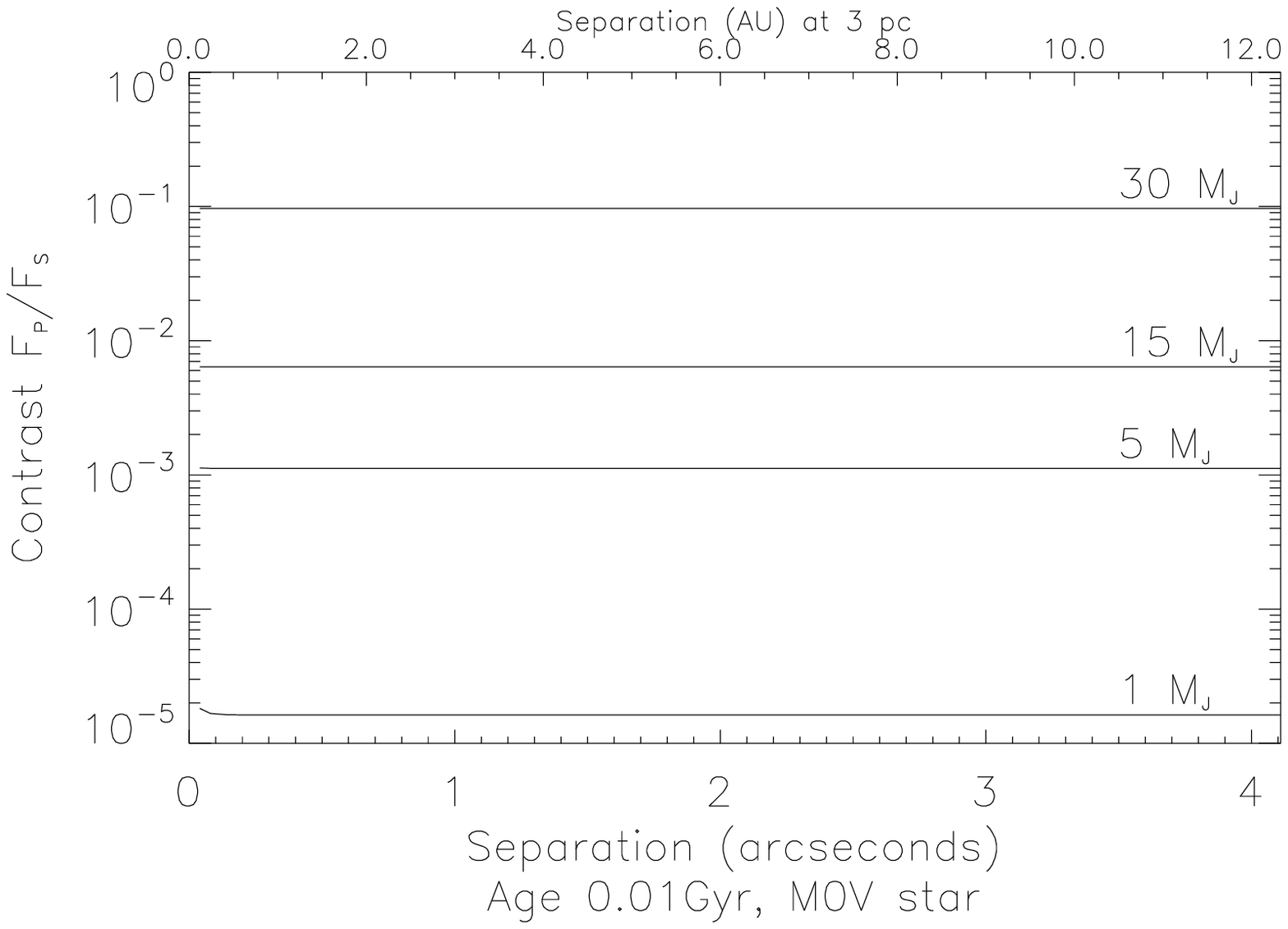}
\plottwo{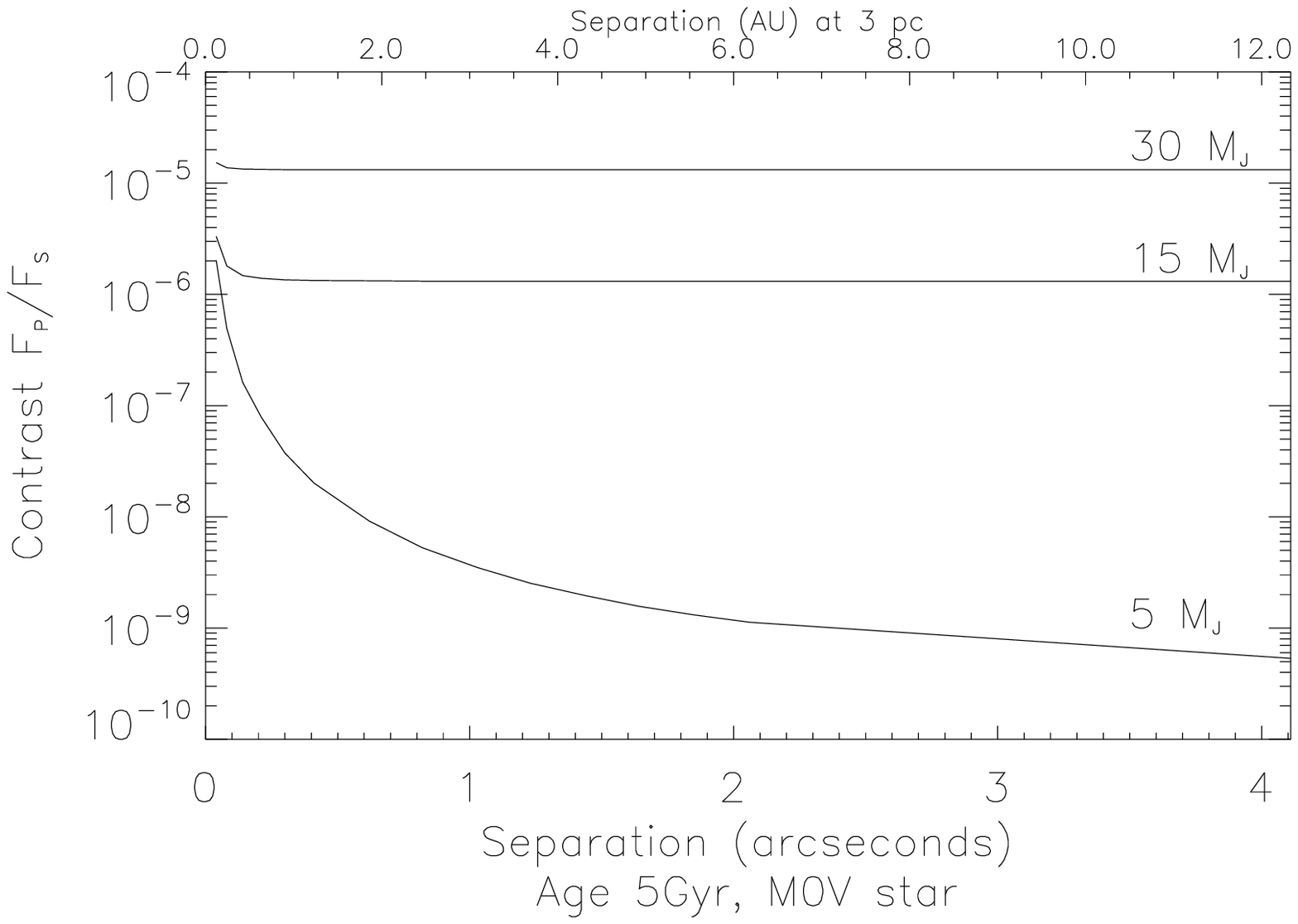}{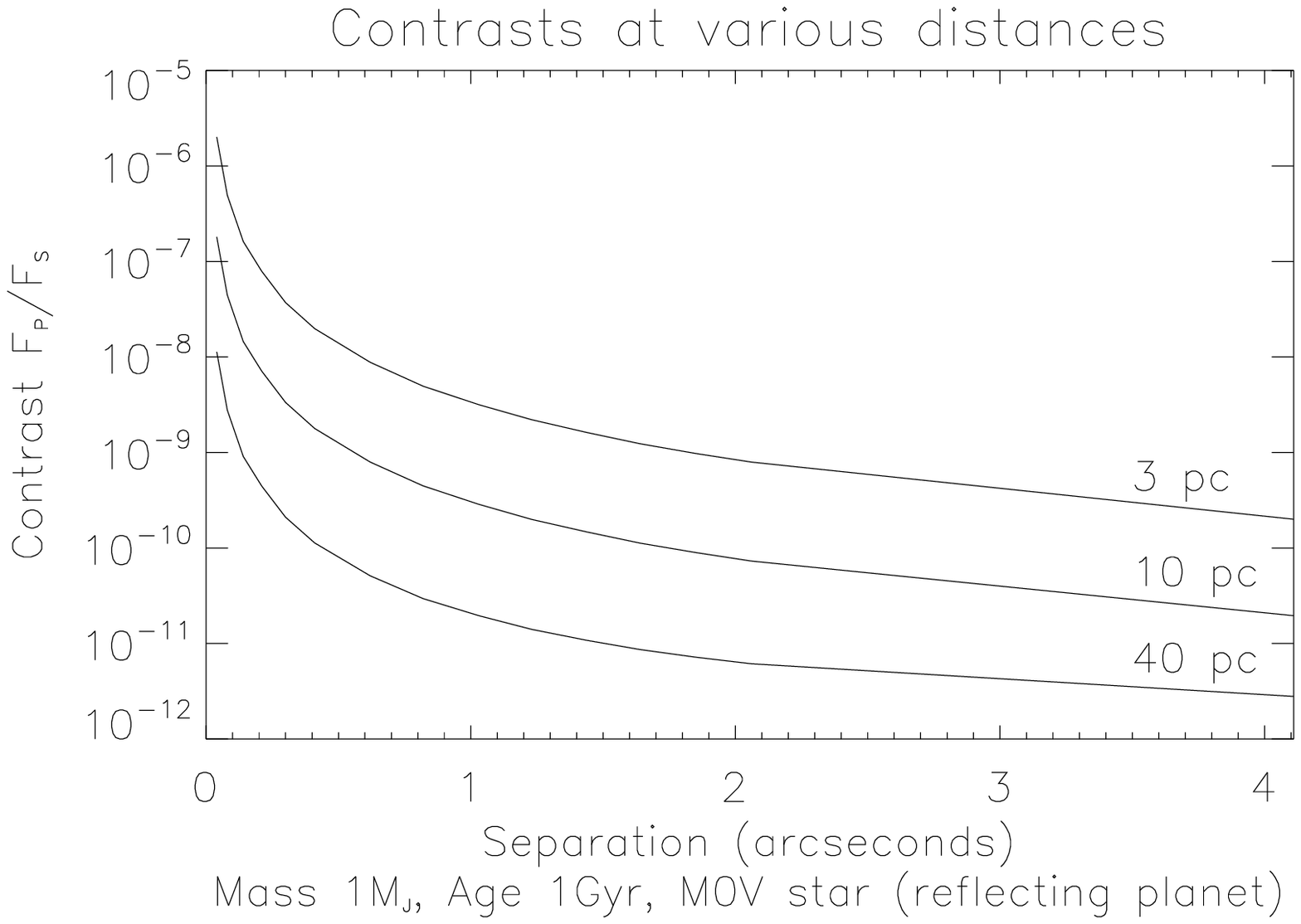}
\caption{Contrasts between the flux of the central peak of a planet ($F_P$) and the
  flux of the central peak of the
  host star ($F_S$) as function of the angular separation, calculated in $J$ band for the planets simulated in
  Sec.\ref{sec:results}. These plots are equivalent to the one in
  Fig.\ref{fig:contrasts}, but the star is a M0V star. Data taken from
  \cite{bar}. See text for details. \label{fig:contrastsM}}
\end{figure}

In Tab.\ref{tab:contrast} and in Fig.\ref{fig:contrasts} and
Fig.\ref{fig:contrastsM} we report an analysis
of the flux contrasts between the intensity peaks of the planets and stars
considered in Sec.\ref{sec:results}, calculated in $J$ band. In the four plots of
Fig.\ref{fig:contrasts} these contrasts have been represented as a function of
the angular separation, always expressed in arcseconds. The
contrast are also plotted with respect to the mass of the planet and to the
age of the system. The contrast are calculated for a distance of 3 pc from the
Sun; moreover, in both Fig.\ref{fig:contrasts} and
Fig.\ref{fig:contrastsM}, at bottom right, the contrasts for a 1
M$_J$ and 1 Gyr old planet at different distances are presented.
The magnitudes of the planets in $J$ band are taken from \cite{bar}, and the
fluxes at various separations are computed used the CSS spreadsheet described
in Appendix \ref{app:css}.

From the plots we can see the different behaviour of the contrast of reflecting planets,
if compared to thermally emitting planets: the formers present a decrement in
flux at large separation, due to the dependance of the flux on the distance
$d^{-2}$ (Eq.\ref{eq:refl}); 
the latters present the same decrement but only very close to the host star,
while at larger separations the reflection of stellar light becomes negligible
with respect to the intrinsic emission, and the contrast becomes a constant.

Even if in the plots the main results are presented for a distance of 3 pc,
the values of contrast obtained for warm thermally emitting planets are
expected to remain the same also at different distances (except for a small
region close to the star). For less massive or older planets this is not true,
because the flux depends on the physical separation between star and planet,
and at a larger distance this separation appears under a smaller angular
separation. For this reason, the plots for different distances have been drawn
only for planet shining by reflecting light (1 M$_J$, 1 Gyr old). 

Tab.\ref{tab:contrast} summarizes these contrasts for all the studied cases,
for planets at a separation of 1 AU from their host star. Since the physical
  separation is fixed, these results are valid at every distance from the
  Sun. 

\section{CHEOPS Simulation Spreadsheet}
\label{app:css}

CSS calculates a set of signal-to-noise ratios of a monochromatic planet image (assumed to
have the size of the Airy disk) at different separations from the star, as
expected as output of a simultaneous differential imager like the CHEOPS
IFS channel. Adapting the instrumental parameters to various cases, it can be used for IFS instruments other than the specific CHEOPS case. The signal is computed in physical
units in different bands ($R$, $I$, $J$, $H$, $K$, $L'$\ and $M-$bands) for
both the star and the planet. Various choices are possible for the basic
parameters: stellar luminosity and distance, various planet ages and masses,
planet-star separation, etc. Both intrinsic emission and reflected stellar
light are considered. The intrinsic emission from planets is estimated using
magnitudes from \citet{bar} COND models. Since a flat spectrum is assumed
within each band, fluxes at the "emission peaks" in the $J$\ and $H$\ bands
are underestimated by a factor of 2-3. For the reflecting light, it is assumed
that the planet is at quadrature, so that the projected physical separation is
the real star-planet distance divided by the distance of the star from us;
and that only a fraction $A_{\lambda}$ of
the light is reflected, $A_{\lambda}$\ being the geometrical albedo from \citet{sud}, dependent on wavelength, and has to be multiplied by the phase factor $\phi$ (0.41 for a specific phase angle of 80$^{\circ}$) to include the 
geometrical effects:
\begin{equation}
\label{eq:refl}
F_{ref} = F_* \frac{ \phi A_{\lambda} R_p^2}{d^2},
\end{equation}

\noindent where $F_{ref}$ is the reflected flux from the planet,  $F_*$ is the stellar flux, $d$ is the distance between star and planet and $R_p$ is the planetary radius. 

Noise estimates include the contributions of the star and planet photon
statistics, sky background, detector, and speckles, this last possibly
suppressed by any predefined value. The photon noise $N_{ph}$ is computed
assuming a Poisson distribution. The sky background ($B$) is computed using values
appropriate for Paranal, in dark sky conditions. Multiple exposures are
explicitly taken into account to estimate the contribution of detector
read-out noise ($RON$). Speckle noise is estimated as the photon noise times the
square root of the total number of photons per speckle. This is obtained by
dividing the number of detected photons (in the selected pass bands) by the
number of speckles $n$, which can be obtained by Eq.\ref{eq:ns}:
\begin{equation}
N_s = N_{ph} \varepsilon \sqrt{\frac{F_*}{n}}.
\end{equation}
The factor of reduction of speckle noise $\varepsilon$, that was in Sec.\ref{sec:epsilon} a function of angular separation, is here an average value for all the separations. Its value, previously calculated to be around 10$^{-3}$, is here enlarged to 10$^{-2}$ to have more conservative results. The total noise $N_t$ can be expressed in this way:
\begin{displaymath}
N_t = \sqrt{N_{ph}^2 + N_s^2 + N_p^2 + B + RON^2} =
\end{displaymath}
\begin{equation}
\quad = \sqrt{f_{seeing}(\theta) \left( 1 + \varepsilon^2 \frac{F_*}{n} \right) + f_{diff}(\theta) + F_p + B + RON^2}.
\end{equation}

We indicated with $f_{seeing}$ and $f_{diff}$ the different contribution given to the total flux by the stellar halo and core, this one being a diffraction-limited PSF. $N_p$, the noise due to the planetary flux, is expressed as the square root of the flux coming from the planet, $F_p$, which is the sum of both reflected and intrinsic emission. 

Various instrumental parameters can be defined, including spectral bandwidth,
instrument efficiency, adaptive optics Strehl ratio, stray light level, etc.
Various options are also possible for the atmospheric conditions (represented
by the coherence time $\tau$, usually adopted to be 3 ms in this paper, and by
the seeing FWHM). The seeing PSF is assumed to be of the form defined by
\citet{rac}. The diffraction PSF is given by a power
law (with exponent $-2.8$), which fits well the PSF produced by detailed
simulations.

\clearpage






\clearpage

\begin{table}
\caption{Main detector parameters introduced in the simulations.}
\label{tab:det}
\centering
\begin{tabular}{l c}
\hline \hline
Parameter & Used value \\
\hline
Read out noise & 10 e$^-$ \\
Gain & 10 e$^- /ADU$ \\
Filling factor & 0.9 \\
Flat field error & 10$^{-4}$ (see Sec.\ref{sec:crosstalk}) \\
Bit resolution & 14 bit \\
Average quantum efficiency & 0.6 \\
Dark current & 0.1 e$^- /s$ \\
Full well & 30\ 000 $ADU$ \\
\hline
\end{tabular}
\end{table}

\clearpage

\begin{table}
\caption{Probability of detection of false alarms, and expected number per final differential image. \label{tab:false}}
\centering
\begin{tabular}{l c c}
\hline \hline
Signal-to-noise & Probability & Number of events \\
\hline
1 & 0.3504 & 1347 \\
2 & 0.0921 & 354 \\
3 & 0.0099 & 38 \\
4 & 4.405$\times$10$^{-4}$ & 1.7 \\
5 & 8.007$\times$10$^{-6}$ & 0.031 \\
6 & 5.974$\times$10$^{-8}$ & 2.30$\times$10$^{-4}$ \\
7 & 1.829$\times$10$^{-10}$ & 7.03$\times$10$^{-7}$ \\
\hline
\end{tabular}
\end{table}

\clearpage

\begin{deluxetable}{cccc ccccccccccc}
\tablecaption{Signal-to-noise ratios of the simulated planets and Brown Dwarfs for seeing 0\farcs6. $ST$ is the spectral type of the star, $d$ the distance in parsecs, $A$ the age in Gyr, $M$ the mass of the planet in M$_J$. The angular separations are expressed in arcseconds. The symbol $>$ indicates a ratio larger than 10$^3$, and it means a sure detection.\label{tab:plotg} }
\startdata
\multicolumn{4}{c}{Angular separations:} & 0.3 & 0.41 & 0.62 & 0.82 & 1.03 & 1.23 & 1.44 & 1.64 & 1.85 & 2.06 & 4.21 \\
$ST$ & $d$ & $A$ & $M$ & \multicolumn{11}{c}{} \\
\hline

G0V &  3 &0.01 & 1 &  29 &  52 & 163 & 427 & 845 &   $>$ &   $>$ &   $>$ &   $>$ &   $>$ &   $>$ \\
G0V &  3 &0.01 & 5 &   $>$ &   $>$ &   $>$ &   $>$ &   $>$ &   $>$ &   $>$ &   $>$ &   $>$ &   $>$ &   $>$ \\
G0V &  3 &0.01 &15 &   $>$ &   $>$ &   $>$ &   $>$ &   $>$ &   $>$ &   $>$ &   $>$ &   $>$ &   $>$ &   $>$ \\
G0V &  3 &0.01 &30 &   $>$ &   $>$ &   $>$ &   $>$ &   $>$ &   $>$ &   $>$ &   $>$ &   $>$ &   $>$ &   $>$ \\
G0V &  3 &   1 & 1 &   1 &   1 &   1 &   1 &   2 &   2 &   2 &   2 &   2 &   2 &   2 \\
G0V &  3 &   1 & 5 &   2 &   3 &   7 &  18 &  34 &  54 &  79 & 105 & 135 & 167 & 634 \\
G0V &  3 &   1 &15 &  42 &  79 & 268 & 802 &   $>$ &   $>$ &   $>$ &   $>$ &   $>$ &   $>$ &   $>$ \\
G0V &  3 &   1 &30 & 442 & 812 &   $>$ &   $>$ &   $>$ &   $>$ &   $>$ &   $>$ &   $>$ &   $>$ &   $>$ \\
G0V &  3 &   5 & 5 &   1 &   1 &   1 &   1 &   2 &   2 &   2 &   2 &   2 &   2 &   3 \\
G0V &  3 &   5 &15 &   3 &   5 &  14 &  37 &  72 & 117 & 170 & 228 & 292 & 364 &   $>$ \\
G0V &  3 &   5 &30 &  23 &  42 & 133 & 351 & 695 &   $>$ &   $>$ &   $>$ &   $>$ &   $>$ &   $>$ \\
        \hline
G0V & 10 &0.01 & 1 &  27 &  50 & 155 & 388 & 726 &   $>$ &   $>$ &   $>$ &   $>$ &   $>$ &   $>$ \\
G0V & 10 &0.01 & 5 &   $>$ &   $>$ &   $>$ &   $>$ &   $>$ &   $>$ &   $>$ &   $>$ &   $>$ &   $>$ &   $>$ \\
G0V & 10 &0.01 &15 &   $>$ &   $>$ &   $>$ &   $>$ &   $>$ &   $>$ &   $>$ &   $>$ &   $>$ &   $>$ &   $>$ \\
G0V & 10 &0.01 &30 &   $>$ &   $>$ &   $>$ &   $>$ &   $>$ &   $>$ &   $>$ &   $>$ &   $>$ &   $>$ &   $>$ \\
G0V & 10 &   1 & 1 &   0 &   0 &   0 &   0 &   0 &   0 &   0 &   0 &   0 &   0 &   0 \\
G0V & 10 &   1 & 5 &   1 &   2 &   5 &  15 &  29 &  44 &  61 &  79 &  99 & 120 & 320 \\
G0V & 10 &   1 &15 &  41 &  78 & 253 & 692 &   $>$ &   $>$ &   $>$ &   $>$ &   $>$ &   $>$ &   $>$ \\
G0V & 10 &   1 &30 & 430 & 771 &   $>$ &   $>$ &   $>$ &   $>$ &   $>$ &   $>$ &   $>$ &   $>$ &   $>$ \\
G0V & 10 &   5 & 5 &   0 &   0 &   0 &   0 &   0 &   0 &   0 &   0 &   0 &   0 &   0 \\
G0V & 10 &   5 &15 &   2 &   4 &  12 &  33 &  61 &  95 & 133 & 174 & 216 & 261 & 650 \\
G0V & 10 &   5 &30 &  23 &  41 & 126 & 318 & 598 & 915 &   $>$ &   $>$ &   $>$ &   $>$ &   $>$ \\
        \hline
G0V & 40 &0.01 & 1 &  24 &  41 & 101 & 196 & 305 & 403 & 486 & 548 & 596 & 631 & 733 \\
G0V & 40 &0.01 & 5 &   $>$ &   $>$ &   $>$ &   $>$ &   $>$ &   $>$ &   $>$ &   $>$ &   $>$ &   $>$ &   $>$ \\
G0V & 40 &0.01 &15 &   $>$ &   $>$ &   $>$ &   $>$ &   $>$ &   $>$ &   $>$ &   $>$ &   $>$ &   $>$ &   $>$ \\
G0V & 40 &0.01 &30 &   $>$ &   $>$ &   $>$ &   $>$ &   $>$ &   $>$ &   $>$ &   $>$ &   $>$ &   $>$ &   $>$ \\
G0V & 40 &   1 & 1 &   0 &   0 &   0 &   0 &   0 &   0 &   0 &   0 &   0 &   0 &   0 \\
G0V & 40 &   1 & 5 &   1 &   1 &   4 &   7 &  12 &  16 &  20 &  23 &  26 &  29 &  35 \\
G0V & 40 &   1 &15 &  35 &  60 & 147 & 286 & 439 & 574 & 681 & 762 & 820 & 864 & 981 \\
G0V & 40 &   1 &30 & 388 & 650 &   $>$ &   $>$ &   $>$ &   $>$ &   $>$ &   $>$ &   $>$ &   $>$ &   $>$ \\
G0V & 40 &   5 & 5 &   0 &   0 &   0 &   0 &   0 &   0 &   0 &   0 &   0 &   0 &   0 \\
G0V & 40 &   5 &15 &   2 &   3 &   8 &  16 &  26 &  35 &  44 &  50 &  57 &  61 &  78 \\
G0V & 40 &   5 &30 &  20 &  34 &  82 & 160 & 250 & 333 & 404 & 458 & 500 & 532 & 626 \\
\hline
M0V &  3 &0.01 & 1 & 286 & 525 &   $>$ &   $>$ &   $>$ &   $>$ &   $>$ &   $>$ &   $>$ &   $>$ &   $>$ \\
M0V &  3 &0.01 & 5 &   $>$ &   $>$ &   $>$ &   $>$ &   $>$ &   $>$ &   $>$ &   $>$ &   $>$ &   $>$ &   $>$ \\
M0V &  3 &0.01 &15 &   $>$ &   $>$ &   $>$ &   $>$ &   $>$ &   $>$ &   $>$ &   $>$ &   $>$ &   $>$ &   $>$ \\
M0V &  3 &0.01 &30 &   $>$ &   $>$ &   $>$ &   $>$ &   $>$ &   $>$ &   $>$ &   $>$ &   $>$ &   $>$ &   $>$ \\
M0V &  3 &   1 & 1 &   1 &   1 &   1 &   1 &   1 &   2 &   2 &   2 &   2 &   2 &   1 \\
M0V &  3 &   1 & 5 &  11 &  20 &  60 & 152 & 288 & 446 & 619 & 793 & 970 &   $>$ &   $>$ \\
M0V &  3 &   1 &15 & 416 & 759 &   $>$ &   $>$ &   $>$ &   $>$ &   $>$ &   $>$ &   $>$ &   $>$ &   $>$ \\
M0V &  3 &   1 &30 &   $>$ &   $>$ &   $>$ &   $>$ &   $>$ &   $>$ &   $>$ &   $>$ &   $>$ &   $>$ &   $>$ \\
M0V &  3 &   5 & 5 &   1 &   1 &   1 &   1 &   2 &   2 &   2 &   2 &   2 &   2 &   3 \\
M0V &  3 &   5 &15 &  24 &  45 & 147 & 409 & 821 &   $>$ &   $>$ &   $>$ &   $>$ &   $>$ &   $>$ \\
M0V &  3 &   5 &30 & 234 & 432 &   $>$ &   $>$ &   $>$ &   $>$ &   $>$ &   $>$ &   $>$ &   $>$ &   $>$ \\
	\hline
M0V & 10 &0.01 & 1 & 262 & 456 &   $>$ &   $>$ &   $>$ &   $>$ &   $>$ &   $>$ &   $>$ &   $>$ &   $>$ \\
M0V & 10 &0.01 & 5 &   $>$ &   $>$ &   $>$ &   $>$ &   $>$ &   $>$ &   $>$ &   $>$ &   $>$ &   $>$ &   $>$ \\
M0V & 10 &0.01 &15 &   $>$ &   $>$ &   $>$ &   $>$ &   $>$ &   $>$ &   $>$ &   $>$ &   $>$ &   $>$ &   $>$ \\
M0V & 10 &0.01 &30 &   $>$ &   $>$ &   $>$ &   $>$ &   $>$ &   $>$ &   $>$ &   $>$ &   $>$ &   $>$ &   $>$ \\
M0V & 10 &   1 & 1 &   0 &   0 &   0 &   0 &   0 &   0 &   0 &   0 &   0 &   0 &   0 \\
M0V & 10 &   1 & 5 &  10 &  16 &  44 &  90 & 146 & 201 & 252 & 296 & 333 & 364 & 469 \\
M0V & 10 &   1 &15 & 377 & 643 &   $>$ &   $>$ &   $>$ &   $>$ &   $>$ &   $>$ &   $>$ &   $>$ &   $>$ \\
M0V & 10 &   1 &30 &   $>$ &   $>$ &   $>$ &   $>$ &   $>$ &   $>$ &   $>$ &   $>$ &   $>$ &   $>$ &   $>$ \\
M0V & 10 &   5 & 5 &   0 &   0 &   0 &   0 &   0 &   0 &   0 &   0 &   0 &   0 &   0 \\
M0V & 10 &   5 &15 &  22 &  38 & 101 & 208 & 333 & 447 & 547 & 626 & 687 & 736 & 884 \\
M0V & 10 &   5 &30 & 212 & 364 & 903 &   $>$ &   $>$ &   $>$ &   $>$ &   $>$ &   $>$ &   $>$ &   $>$ \\
	\hline
M0V & 40 &0.01 & 1 & 139 & 199 & 355 & 524 & 628 & 681 & 707 & 722 & 730 & 736 & 748 \\
M0V & 40 &0.01 & 5 &   $>$ &   $>$ &   $>$ &   $>$ &   $>$ &   $>$ &   $>$ &   $>$ &   $>$ &   $>$ &   $>$ \\
M0V & 40 &0.01 &15 &   $>$ &   $>$ &   $>$ &   $>$ &   $>$ &   $>$ &   $>$ &   $>$ &   $>$ &   $>$ &   $>$ \\
M0V & 40 &0.01 &30 &   $>$ &   $>$ &   $>$ &   $>$ &   $>$ &   $>$ &   $>$ &   $>$ &   $>$ &   $>$ &   $>$ \\
M0V & 40 &   1 & 1 &   0 &   0 &   0 &   0 &   0 &   0 &   0 &   0 &   0 &   0 &   0 \\
M0V & 40 &   1 & 5 &   5 &   7 &  14 &  22 &  29 &  31 &  34 &  34 &  35 &  35 &  37 \\
M0V & 40 &   1 &15 & 203 & 288 & 507 & 729 & 860 & 921 & 952 & 968 & 978 & 985 & 997 \\
M0V & 40 &   1 &30 &   $>$ &   $>$ &   $>$ &   $>$ &   $>$ &   $>$ &   $>$ &   $>$ &   $>$ &   $>$ &   $>$ \\
M0V & 40 &   5 & 5 &   0 &   0 &   0 &   0 &   0 &   0 &   0 &   0 &   0 &   0 &   0 \\
M0V & 40 &   5 &15 &  11 &  16 &  30 &  48 &  61 &  68 &  73 &  75 &  76 &  78 &  80 \\
M0V & 40 &   5 &30 & 113 & 162 & 292 & 435 & 529 & 577 & 601 & 615 & 623 & 628 & 639 \\
\hline
\enddata
\end{deluxetable}

\clearpage

\begin{table}
\caption{Flux contrasts between the intensity peaks of the planets and stars
  considered in Tab.\ref{tab:plotg}, for a fixed separation of 1 AU between
  the two objects. For planets characterized by intrinsec thermal emission we
  don't expect a significant decrement of their fluxes for larger separations;
  so the contrasts calculated for 1 AU can be assumed to be valid also further
  away from the star. This is not true for planets shining by reflected light,
  whose brightness strongly depends on the physical separation (see e.g. the 1
  M$_J$ planet in Fig.\ref{fig:contrasts} at top-left). \label{tab:contrast}}  
\centering
\begin{tabular}{l l c c c c}
\hline \hline
 & Mass: & 1M$_J$ & 5M$_J$ & 15M$_J$ & 30 M$_J$ \\
St. Type & Age & & & & \\ 
\hline
G0V & 0.01Gyr & 1.60$\cdot$10$^{-6}$ & 1.08$\cdot$10$^{-4}$ &
 6.16$\cdot$10$^{-4}$ & 9.33$\cdot$10$^{-3}$ \\
G0V & 1Gyr & 3.71$\cdot$10$^{-8}$ & 9.41$\cdot$10$^{-8}$ &
 2.34$\cdot$10$^{-6}$ & 2.52$\cdot$10$^{-5}$ \\
G0V & 5Gyr & - & 3.71$\cdot$10$^{-8}$ &
 1.63$\cdot$10$^{-7}$ & 1.31$\cdot$10$^{-6}$ \\
M0V & 0.01Gyr & 1.62$\cdot$10$^{-5}$ & 1.12$\cdot$10$^{-3}$ &
 6.39$\cdot$10$^{-3}$ & 9.68$\cdot$10$^{-2}$ \\
M0V & 1Gyr & 3.71$\cdot$10$^{-8}$ & 6.40$\cdot$10$^{-7}$ &
 2.39$\cdot$10$^{-5}$ & 2.62$\cdot$10$^{-4}$ \\
M0V & 5Gyr & - & 3.77$\cdot$10$^{-8}$ &
 1.35$\cdot$10$^{-6}$ & 1.32$\cdot$10$^{-5}$ \\
\hline
\end{tabular}
\end{table}







\end{document}